\documentclass[fleqn,usenatbib]{mnras}

\usepackage{newtxtext,newtxmath}
\usepackage{threeparttable}

\usepackage[T1]{fontenc}
\usepackage{booktabs}
\DeclareRobustCommand{\VAN}[3]{#2}
\let\VANthebibliography\thebibliography
\def\thebibliography{\DeclareRobustCommand{\VAN}[3]{##3}\VANthebibliography}

\usepackage{graphicx}	
\usepackage{amsmath}	

\title[Spinning Dust Emission in Barnard~30]{COMAP Galactic Science I: Observations of Spinning Dust Emission at 30\,GHz in Dark Clouds Surrounding the $\lambda$-Orionis H\textsc{ii} Region}

\author[S. E. Harper et al.]{Stuart E. Harper,$\!^{1}$\thanks{E-mail: stuart.harper@manchester.ac.uk}, Clive Dickinson,$\!^{1}$ Kieran A.~Cleary,$\!^{2}$ Brandon S.~Hensley,$\!^{3}$ \newauthor Gabriel A.~Hoerning,$\!^{1}$ Roberta Paladini,$\!^{4}$   Thomas J. Rennie,$\!^{1,5}$ R.\,Cepeda-Arroita,$\!^{1}$ \newauthor Delaney A.~Dunne,$\!^{2}$   Hans Kristian Eriksen,$\!^{6}$ Joshua Ott~Gundersen,$\!^{7}$ H\aa vard T.~Ihle,$\!^{6}$  \newauthor  Jonas G. S. Lunde,$\!^{6}$Roberto Ricci,$\!^{8}$   Jeroen Stil,$\!^{9}$ Nils-Ole~Stutzer,$\!^{6}$ A. R.~Taylor,$\!^{10,11,12}$  \newauthor  Ingunn Kathrine Wehus$^{6}$
\\
$^1$Jodrell Bank Centre for Astrophysics, Department of Physics \& Astronomy, The University of Manchester, \\
Oxford Road, Manchester, M13 9PL, U.K. \\ 
$^2$California Institute of Technology, 1200 E. California Blvd., Pasadena, CA 91125, USA \\
$^3$Jet Propulsion Laboratory, California Institute of Technology, 4800 Oak Grove Drive, Pasadena, CA 91109, USA \\
$^4$Infrared Processing Analysis Center, California Institute of Technology, Pasadena, CA 91125, USA \\
$^5$Department of Physics and Astronomy, University of British Columbia, Vancouver, BC Canada V6T 1Z1, Canada \\
$^6$Institute of Theoretical Astrophysics, University of Oslo, P.O. Box 1029 Blindern, N-0315 Oslo, Norway \\
$^7$Department of Physics, University of Miami, 1320 Campo Sano Avenue, Coral Gables, FL 33156 \\
$^8$Instituto di Radioastronomia, Bologna, Via Gobetti 101, I-40129, Italy \\
$^9$Department of Physics and Astronomy, University of Calgary, Alberta, Canada \\
$^{10}$Inter-University Institute for Data Intensive Astronomy, Cape Town, South Africa \\
$^{11}$Department of Astronomy, University of Cape Town, South Africa \\
$^{12}$Department of Physics and Astronomy, University of the Western Cape, South Africa
}

\date{Accepted XXX. Received YYY; in original form ZZZ}

\pubyear{2024}

\begin{document}
\label{firstpage}
\pagerange{\pageref{firstpage}--\pageref{lastpage}}
\maketitle

\begin{abstract}
Anomalous Microwave Emission (AME) is a major component of Galactic emission in the frequency band 10--60\,GHz and is commonly modelled as rapidly rotating spinning dust grains. The photodissociation region (PDR) at the boundary of the $\lambda$-Orionis H\textsc{ii} region has been identified by several recent analyses as one of the brightest spinning dust emitting sources in the sky. We investigate the Barnard~30 dark cloud, a dark cloud embedded within the $\lambda$-Orionis PDR. We use total-power observations of Barnard~30 from the CO Mapping Array Project (COMAP) pathfinder instrument at 26--34GHz with a resolution of $4.\!^\prime5$ alongside existing data from \textit{Planck}, WISE, IRAS, ACT, and the 1.447\,GHz GALFACTS survey. We use aperture photometry and template fitting to measure the spectral energy distribution of Barnard~30. We find that the spinning dust is the dominant emission component in the 26--34GHz range at the $7\,\sigma$ level ($S_{30\mathrm{GHz}} = 2.85\pm0.43$\,Jy). We find no evidence that polycyclic aromatic hydrocarbons are the preferred carrier for the spinning dust emission, suggesting that the spinning dust carriers are due to a mixed population of very small grains. Finally, we find evidence for variations in spinning dust emissivity and peak frequency within Barnard~30, and that these variations are possibly driven by changes in dust grain population and the total radiation field. Confirming the origin of the variations in the spinning dust spectrum will require both future COMAP observations at 15\,GHz combined with spectroscopic mid-infrared data of Barnard~30.
\end{abstract}




\section{Introduction}

Anomalous microwave emission (AME) refers to an excess emission at microwave frequencies that correlates with thermal dust emission at higher frequencies \citep[see][for a review]{Dickinson2018}. It was first detected several decades ago as a new foreground by CMB experiments \citep{Kogut1996, Leitch1997}. Since then, AME has been detected from a range of environments, including dark clouds \citep[e.g.,][]{Casassus2006, Harper2015, Vidal2019}, H\textsc{ii} regions \citep{Dickinson2007, Tibbs2012, Rennie2022}, cirrus dust clouds \citep{Hensley2016, Harper2022}, photo-dissociation regions (PDRs) \citep{Casassus2008}, and many other regions within the Galaxy \citep{PIP_XV, Poidevin2023}. AME has also had several detections in other galaxies \citep{Murphy2010,Hensley2015,Harper2024}.  The leading model suggests that AME is generated by the rapid rotation of very small dust grains, proposed as the spinning dust model by \citet{Draine1998a, Draine1999}, and later refined by the publicly available \textsc{SpDust} code \citep{Ali-Hamoud2009, Silsbee2011}. The exact carrier for spinning dust emission is still unknown. Polycyclic aromatic hydrocarbons (PAHs) were initially favored as a carrier due to their prevalence in the interstellar medium (ISM) and being of the correct physical size needed to produce the observed spectrum \citep{Draine1998a}. However, there is no strong observational evidence for PAHs being the favored carrier, implying that spinning dust could be generated by a continuum of very small dust grains (VSGs), nanograins \citep{Hoang2016,Hensley2016,Ysard2022}, or nanodiamonds \citep{Greaves2018}, which could be both carbonaceous and silicate in nature.

$\lambda-$Orionis is an O8 III type star in the Collinder-69 cluster that is surrounded by a well-studied H\textsc{ii} region \citep{Sharpless1959} characterised by a bright radio/optical ionised region at the center surrounded by a prominent IR photodissociation region (PDR) \citep{Maddalena1987, Lang2000}. The $\lambda$-Orionis association, located approximately 400\,pc away, is thought to have formed around 5\,Myr ago \citep{Murdin1977}. The angular diameter of the $\lambda$-Orionis H\textsc{ii} region is approximately $5^\circ$, making it an ideal target for several low-resolution spinning dust emission studies \citep{PIP_XV, Bell2019, CepedaArroita2021,Chuss2022} that found spinning dust to be the dominant emission mechanism around 30\,GHz. The question of the carrier for spinning dust in this region remains open. PAHs were preferred by the Bayesian analysis conducted by \citet{Bell2019}, while correlations between maps of spinning dust emission derived from the \textit{Planck} \textsc{commander} analysis \citep{Planck2015_X} and a map of the $3.3$\,$\mu$m PAH feature from COBE/DIRBE data disfavored PAHs \citep{Chuss2022}. However, these studies were limited to observations at 1\,degree scales, often masking out the brightest spinning dust regions within the PDR. Fully characterizing the origin of spinning dust emission within the $\lambda$-Orionis PDR requires arcminute-resolution observations that can resolve individual clouds.

\begin{figure}
    \includegraphics[width=\columnwidth]{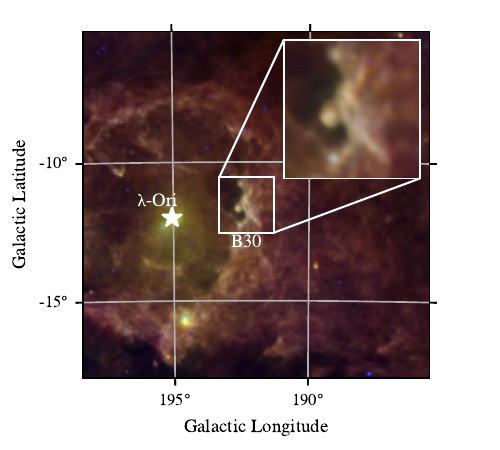}
    \caption{Location of the Barnard~30 (B30) dark cloud within the $\lambda$-Orionis PDR. The image is a three-colour composite of the IRAS 100\,$\mu$m (red), 60\,$\mu$m (green), and 25\,$\mu$m (blue) data, smoothed to $6^\prime$ resolution. The location of the $\lambda$-Orionis star is indicated by the \textit{white} star. The inset image shows a zoomed in view of the B30 region that was observed by COMAP. The image has dimensions of $12.\!\!^\circ5 \times 12.\!\!^\circ5$. The colour map of the image has had an arcsinh stretch applied.}
    \label{fig:iras_finder}
\end{figure}

We use dedicated observations by the CO Mapping Array Project (COMAP) pathfinder instrument to observe a section of the PDR around $\lambda$-Orionis at a resolution of $4.\!^\prime5$, between 26 and 34\,GHz. The observations are centred upon the dark cloud Barnard~30 \citep[B30; ][]{Dolan2001,Koenig2015}. \autoref{fig:iras_finder} indicates the location of B30 within the $\lambda$-Orionis ring as seen in the IRAS 100\,$\mu$m map, which is a good tracer of warm $(T\sim 20$\,K) dust. The B30 region was first proposed as a spinning dust emission candidate by \textit{Planck} \citep{PIP_XV}, and later confirmed by \cite{CepedaArroita2021} when including 5\,GHz C-BASS data. B30 is a bright-rimmed cloud with a PDR facing towards the $\lambda$-Orionis star. Numerous AME observations have identified PDRs as strong sources of spinning dust emission \citep[e.g.,][]{Casassus2008}, and the strong environmental variations throughout a PDR make them an excellent laboratory for the study of variations in spinning dust emissivity with dust grain population using high resolution radio observations. 

This paper is structured as follows: In \autoref{sec:data} we describe the COMAP survey and all ancillary data used; in \autoref{sec:photometry} we describe the aperture photometry method used to measure the integrated flux density of the cloud; in \autoref{sec:template-fitting} we describe how template fitting is used to decompose B30; and in \autoref{sec:conclusions} we present our conclusions.

\section{Data}\label{sec:data}

In the following sections we describe the COMAP instrument and observations, and all of the ancillary data used for this analysis. All data are smoothed to a common resolution of $6^\prime$ full-width half-maximum (FWHM).

\subsection{COMAP} 

The COMAP instrument is a spectroscopic radiometer with a spectral resolution of $\sim2$\,MHz between 26 and 34\,GHz. The front-end of the instrument consists of 19 closely-packed left-hand polarised circular feed horns. It is fielded on a 10.4\,m Cassegrain design telescope based at the Owens Valley Radio Observatory (OVRO) that was previously used as part of the CARMA experiment \citep{Leighton1977}. The telescope has a FWHM of $4.\!^\prime5$ at 30\,GHz, with optics designed to ensure that the FWHM varies by less than $\pm 4$\,per\,cent across the COMAP band. Each feed horn is separated on the sky by $12.\!^\prime04$, resulting in 19 simultaneous and independent observations of the sky.

 A single circular polarization of the incoming radiation is amplified by a cooled low-noise amplifier, then down-converted at ambient temperature before being split into two 4\,GHz bands. These 4\,GHz bands are then digitised and passed to two ROACH-2 spectrometers \citep{Parsons2008} which perform sideband separation. For each feed, we obtain four 2\,GHz wide sidebands, sampled once every 20\,ms. More details on the COMAP instrument can be found in \citep{Lamb2022}.

We generally employ the same data reduction methods used in the first COMAP Galactic science release \citep{Rennie2022}. Here, we provide a brief overview of the data analysis pipeline and any modifications made. The COMAP data are initially calibrated by covering the focal plane with a microwave-absorbing vane of known temperature at the beginning and end of each observation ($\sim 1$\,hour). Subsequently, the data are calibrated onto the astronomical brightness temperature scale using daily measurements of the supernova remnants Taurus~A (Tau~A) and Cassiopeia~A (Cas~A), along with the brightness models supplied by the WMAP 7-year release \citep{Weiland2011}. The overall calibration uncertainty is $3.2 \pm 0.1$\,per\,cent. 

Following the approach of \citet{Rennie2022}, we subtract the mean atmospheric brightness using an atmospheric slab model and apply a running median filter to the data. In contrast to the main Galactic science analysis, we have also implemented a novel continuous-gain correction algorithm capable of mitigating $1/f$ noise fluctuations by fitting system temperature templates to each time sample. This method enhances the sensitivity of a substantial portion of the COMAP continuum data by a factor of two or more. We defer detailed discussion of this algorithm to a future COMAP publication (Lunde et al., in prep). The COMAP spectroscopic data for this analysis are averaged into four 2\,GHz bands with central frequencies of 27, 29, 31, and 33\,GHz. Lastly, the final maps are created using an implementation of the destriping map-making algorithm \citep[e.g.,][]{Delabrouille1998, Sutton2010}, which is the same implementation used for the Galactic science data \citep{Rennie2022}.

\subsection{COMAP Observations of Barnard 30}

\begin{figure}
    \centering
    \includegraphics[width=\columnwidth]{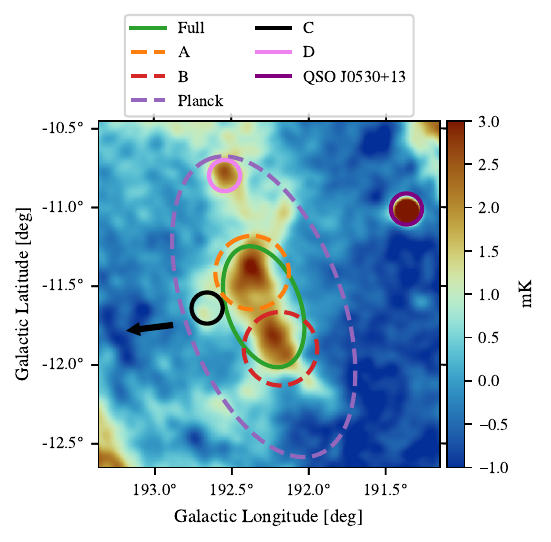}
    \caption{Overview of the target B30 cloud at the edge of the $\lambda$-Orionis H\textsc{ii} region as seen in the COMAP 27\,GHz data, smoothed to $6^\prime$. The green region (\textit{Full}) indicates the region defined as the main cloud as seen in COMAP, while regions A and B indicate the smaller sub-region hotspots, respectively. Regions C and D are smaller nearby dust clouds that do not share the same spectrum as the bulk of the cloud, these are discussed in \autoref{sec:photometry}. We also mark the location of \mbox{QSO~J030+13}, a QSO within the field and which is discussed in \autoref{sec:qso}. The black arrow indicates the direction of the central ionizing $\lambda$-Orionis star.}
    \label{fig:finder}
\end{figure}

The observed field was a $3 \times 3$\,deg$^2$ region centred on LDN\,1582\footnote{R.A. and Declination of 05:32:00.3, +12:30:28}, a dark cloud within Barnard 30. For the purposes of this analysis, when referring to B30 we mean the entire region indicated within the inset image of \autoref{fig:iras_finder}, which we break up into several sub-regions. References to $\lambda$-Orionis refer to both the stellar association and the larger $\lambda$-Orionis PDR within which B30 is embedded. Observations were taken between May 2020 and December 2022. The field was observed for approximately 270\,hours. After removing bad observations due to the proximity of the Sun, bad weather, or instrumental issues we were left with a total 200\,hours of good data. We used the same Lissajous scanning strategy as for the Galactic plane observations; more details on the scanning strategy and its benefits are given in \citep{Rennie2022}.

\autoref{fig:finder} shows the COMAP 27\,GHz data smoothed to $6^\prime$ to reduce background noise. The field covers the entire B30 cloud plus one nearby radio source, QSO~J030+13 (see \autoref{sec:qso} for the measured flux densities of this source at 26--34\,GHz). The black arrow indicates the direction of the $\lambda$-Orionis O8 III type star in Collinder-69 cluster. The apertures indicate the sub-regions that were used to divide up the data, the details of which are discussed in \autoref{sec:photometry}.

The white-noise sensitivity of the COMAP maps is approximately 50$\,\mu$K/beam ($\approx 2$--3\,mJy/beam) at a resolution of $4.\!^\prime5$ in all the 27 to 33\,GHz maps. However, the noise is dominated primarily by large-scale systematics errors in the data that are mitigated using destriping map-making and a map-based median filter with a kernel size of $1^\circ$ (much larger than the B30 cloud). In general we will use the COMAP 27\,GHz band as the main dataset for this analysis as the peak brightness temperature of the signal at 27\,GHz  is approximately 50\,per\,cent higher than in the 33\,GHz band.


\subsection{Ancillary Data} 

We utilise the \textit{Planck} satellite mission PR3 data release \citep{Planck2018_I}, which can be accessed via the Planck Legacy Archive\footnote{\url{http://pla.esac.esa.int/pla}}. We use the \textit{Planck} bands between 217 and 857\,GHz, all smoothed to a common resolution of $6^\prime$ to match the smoothed COMAP data. We do not use the LFI bands, specifically the \textit{Planck} 28.4\,GHz data, as the resolution of this data set is limited to $32.\!^\prime4$---too coarse to make a direct comparison with the COMAP data despite the frequency overlap. All the \textit{Planck} maps were reprojected from a \textsc{HEALPix} grid \citep{Gorski2005} to a regular Cartesian grid. 

We traced the ionised gas associated with $\lambda$-Orionis using both radio and H$\alpha$ data. The radio data comes from the Galactic Arecibo L-band Feed Array Continuum Transit Survey (GALFACTS), which was an L-band transit survey taken between December 2008 and September 2015 using the ALFA receiver \citep{Taylor2013}. We used band-averaged GALFACTS data that have an effective central frequency of 1447\,MHz and a resolution of $3.\!^\prime35$. We also subtract from the GALFACTS map all of the point sources within the field found in the NRAO VLA Sky Survey (NVSS)  catalogue \citep{Condon1998}. We subtract the background radio sources as we only want to use the GALFACTS data as a tracer of the radio free-free emission at the COMAP frequencies.

The H$\alpha$ data come from the Virginia Tech Spectral-Line Survey (VTSS) \citep{Dennison1998}. H$\alpha$ emission is a tracer of ionised gas, and can be directly related to the brightness of the radio free-free continuum emission \citep{Draine_book}. We do not directly use the H$\alpha$ data in this analysis due to significant dust extinction, but they are used to verify the observed GALFACTS radio brightness. The H$\alpha$ data have a resolution of $1.\!^\prime6$.

\begin{figure*}
    \centering
    \includegraphics[width=\textwidth]{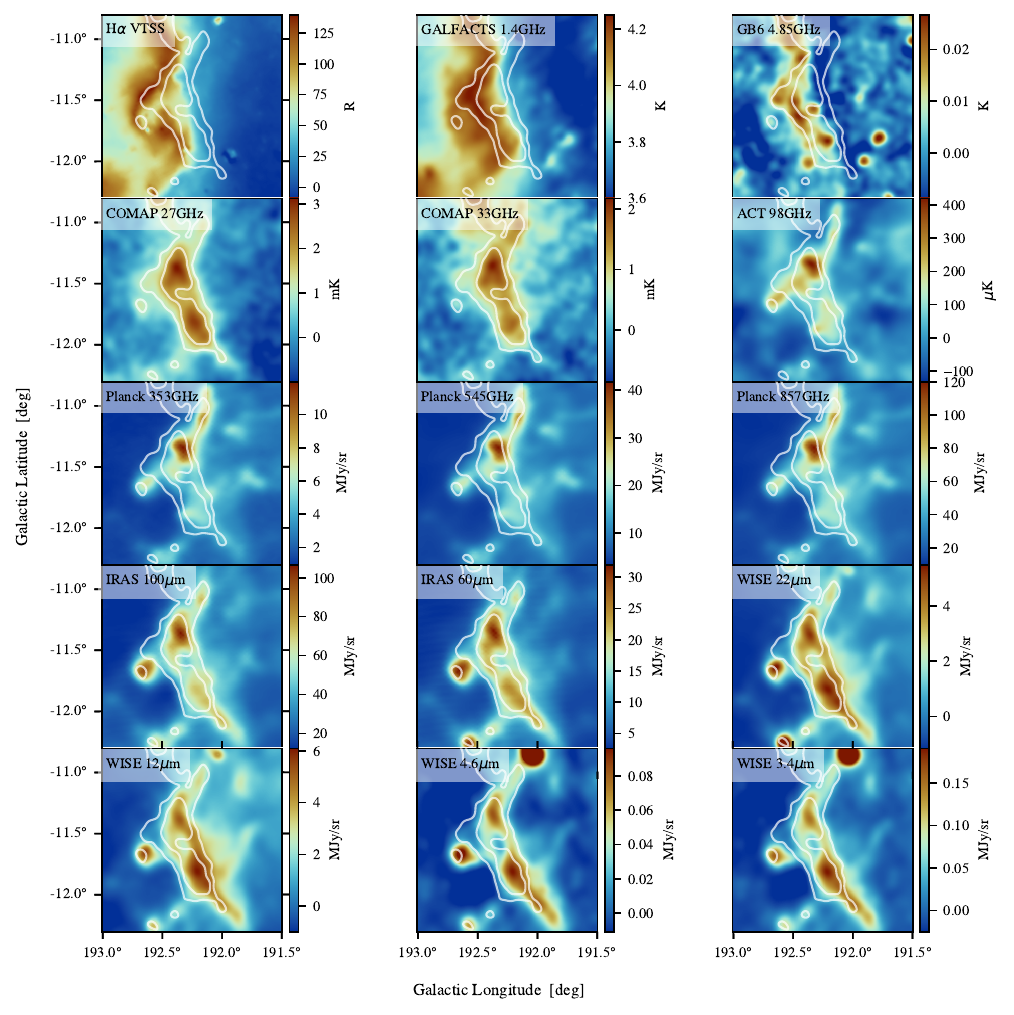}
    \caption{Grid of multi-frequency maps centred on Barnard~30 in ascending order of frequency from the top-left to the bottom-right. All maps have been smoothed to a common resolution of $6^\prime$ except for the VTSS and GB6 maps, which are shown at the native resolution of the each survey ($1.\!^\prime6$ and $3.\!^\prime5$, respectively). The white contours are the COMAP 27\,GHz data at levels of 0.5 and 1.6\,mK. The width and height of each panel is $1.\!\!^\circ$5.}
    \label{fig:grid}
\end{figure*}

We also use data from the Green Bank 4.85\,GHz Northern sky survey \citep[GB6,][]{Gregory1996} to verify the morphology of the free-free emission seen at 1.447\,GHz by GALFACTS. Unfortunately, we cannot directly use the GB6 data in the analysis due to heavy filtering of the large-scale features (evidenced by the negatives around the PDR front shown in \autoref{fig:grid}). The GB6 data were taken from the NASA Skyview website\footnote{\url{https://skyview.gsfc.nasa.gov/}}. The GB6 data have a resolution of $3.\!^\prime5$. 

We used the Atacama Cosmology Telescope (ACT) 98\,GHz intensity data from the night-time subset of ACT DR5 \citep[dr5.01,][]{Naess2020}, available on the NASA Lambda repository\footnote{\url{https://lambda.gsfc.nasa.gov/product/act/actpol_prod_table.html}}. We use the ACT 98\,GHz data in place of the \textit{Planck} 143\,GHz data as the former has a resolution of $2^\prime$, and the \textit{Planck} data have a resolution of $7.\!\!^\prime3$.

We used the four reprocessed Infrared Astronomical Satellite  \citep[IRAS,][]{Neugebauer1984} data \citep[IRIS,][]{Miville-Deschenes2005} maps at $12$, $25$, $60$, and $100$\,$\mu$m. The spatial resolution of IRAS is $\approx 4^\prime$. Here we use the IRAS datasets to constrain the turn over in the thermal dust spectrum that is not measured by the \textit{Planck} data. 

We used all four bands ($3.4$, $4.6$, $12$ and $22$\,$\mu$m) from the Wide-field Infrared Survey Explorer \citep[WISE,][]{Wright2010WISE}. Within the $3.4$ and $12$\,$\mu$m bands there are multiple prominent PAH features \citep[e.g.][]{Tielens2008}, which are a leading candidate for the carrier for spinning dust emission \citep{Dickinson2018}. While the $4.6$ and $22$\,$\mu$m WISE bands contain no bright PAH features and so can be used to trace small grain populations. We converted all of the maps from detector units to MJy/sr using the procedure outlined in \citet{Cutri2012}, and applied a $1^\prime$ rolling 2D median filter to remove contamination of the diffuse emission due to stellar sources in the $3.4$ and $4.6$\,$\mu$m data.

\autoref{fig:grid} shows all the ancillary data alongside the 27 and 33\,GHz COMAP data all smoothed to $6^\prime$. We have overlaid the COMAP 27\,GHz contours over each image for easier comparison. The H$\alpha$, GALFACTS and GB6 maps are placed in the first two panels as these are tracers of the ionised gas component being driven by the $\lambda$-Orionis star. We have not corrected the VTSS data for dust extinction, which is why the GALFACTS and VTSS data are not identical. All other images increase in frequency from top-left to bottom-right.

\section{Aperture Photometry}\label{sec:photometry} 

\begin{table}
    \caption{Galactic coordinates and dimensions used for each aperture shown in \autoref{fig:finder}. Note that $\theta$ is given relative to Galactic North. } \label{tab:apertures}
    \begin{tabular}{cccccc}
    \hline
        Region Name & $\ell$ & $b$ & semi-major & semi-minor & $\theta$ \\
                    & deg.    & deg. & arcmin.   & arcmin.   & deg. \\ \hline 
        \textit{Full}        & 192.29  & $-11.63$ & 24      & 14        & $-20$ \\ 
        A           & 192.03  & $-11.42$ & 14      & 14        & 0 \\ 
        B           & 192.18  & $-11.90$ & 14      & 14        & 0 \\ 
        C           & 192.60  & $-11.67$ & 14      & 14        & 0 \\ 
        D           & 192.53  & $-10.76$ & 14      & 14        & 0 \\ 
        \textit{Planck}      & 192.29  & $-11.63$ & 60      & 30        & $-20$ \\     \midrule 
        QSO~J0530+11 & 191.36 & -11.03 & 14 & 14 & 0 \\
        \hline      
    \end{tabular}
\end{table}

\begin{table}
    \caption{Background aperture descriptions. The central coordinate of each group of background apertures is given in \autoref{tab:apertures}. $N$ refers to the number of background apertures used to sample the annulus. The final column, $\alpha_c$, gives the maximum aperture correction for each configuration. See text for details. } \label{tab:background}
    \begin{tabular}{ccccccc}
    \hline
        Region Name & N & radius & semi-major & semi-minor & $\alpha_c$ \\
                    &     & arcmin. & arcmin.   & arcmin.   &   \\ \hline 
        \textit{Full}        &  11   &  8    & 34      & 24        & 1.11 \\ 
        A           &  9  &   8   & 24      & 24        &  1.14 \\ 
        B           &  9  &   8   & 24      & 24        &  1.14 \\ 
        C           &  9  &   8   & 24      & 24        &  1.14 \\ 
        D           &  9  &   8   & 24      & 24        &  1.14 \\ 
        \textit{Planck}  & 17 & 10 & 70      & 40        &  1.05 \\     \midrule 
        QSO~J0530+11 & 9 & 8 & 24 & 24 &  1.14 \\
        \hline      
    \end{tabular}
\end{table}

To measure the integrated flux density $S_\nu$ of the cloud we used aperture photometry. Aperture photometry is defined as 
\begin{equation}
    S_\nu = 2k_b \frac{\nu^2}{c^2}\int_{\Omega} \left[T(\Omega^\prime) - \left<T_\mathrm{bkgd}\right> \right] \mathrm{d} \Omega^\prime,
\end{equation}
where $\Omega$ is the area of the aperture enclosing the source, $T$ is the source brightness distribution, and $\left<T_\mathrm{bkgd}\right>$ is the average background brightness sampled from a low-brightness region around the source. In this analysis we used ellipses, defined in \autoref{tab:apertures}, to measure source flux density in each region shown in \autoref{fig:finder}. Whereas the background brightness was sampled from $N_\mathrm{bkgd}$ independent circular apertures that were equally spaced along an ellipse that enclosed the main aperture. \autoref{tab:background} defines the path of the background apertures, the radii of each, and the number ($N$) of background apertures used for each source. 

The advantage of calculating the background using many independent apertures instead of a single annulus is that the background of the source could be sampled with $N_\mathrm{bkgd} (N_\mathrm{bkgd} -1)$ independent configurations. For each configuration we measured the source background brightness using both the median brightness and a fitted 2D plane. This resulted in flux density distributions using all median and plane fitted backgrounds brightness configurations. We found that, except for the 1.4\,GHz radio data, there was no significant (within uncertainties) difference between the median flux density of both background distributions. As such, we use just the median flux density of the median background brightness distributions.  

The standard method for measuring the uncertainty in the flux density for a source with a white noise background is
\begin{equation}\label{eqn:uncertainties}
\sigma_S^2 = \sigma_\mathrm{annu}^2 N_S \left[1 + \frac{\pi}{2}\frac{N_S}{N_\mathrm{annu}}\right] + \delta^2 S^2
\end{equation}
where, $\delta$ represents the calibration uncertainty, $\sigma_\mathrm{annu}$ is the standard deviation of the annulus pixels in units of Jy/pixel, $N_S$ is the number of pixels within the aperture, and $N_\mathrm{annu}$ is the number of background pixels. However, in this case the background noise in all of the maps was neither Gaussian nor white. As such we used numerical simulations to calculate the uncertainty in the flux density by injecting sources of known flux density at 100 random locations within each map, and measuring the flux density using the same aperture used in the analysis. In this way we could measure the uncertainty in the flux density by measuring the standard deviation of the resulting distribution. The numerically derived uncertainties were found to be consistent with \autoref{eqn:uncertainties} in white noise only simulations. We find that for COMAP, the numerically derived uncertainties are approximately 10 times higher than those predicted from the white-noise alone due to a combination of atmospheric and large-scale noise features in the map. For other datasets, such as WMAP and \textit{Planck}, we find that the numerically derived uncertainty is approximately twice what would be expected from white-noise alone, this difference mostly being driven by variations in the background emission around B30. 

Finally, \autoref{tab:background} reports the aperture corrections applied to each aperture. Aperture corrections account for the amount of flux density from the source that is measured in the background pixels and is then subtracted from the source flux density. Aperture corrections are a function of the beam used to smooth the sky signal and the separation of the background pixels from the aperture pixels. To calculate this we made a mock dataset with the aperture pixels set to unity and all other pixels set to zero. We smoothed the mock data to the resolution of $6^\prime$ and then measured the flux density within the aperture as we would in the real data. The aperture correction is then defined as 
\begin{equation}
    \alpha_c = \frac{S_\mathrm{full}}{S_\mathrm{measured}}, 
\end{equation}
where $S_\mathrm{full}$ is equal to the input flux density within the aperture, and $S_\mathrm{measured}$ is the flux density measured using our aperture photometry technique. The reported flux densities are scaled by the coefficient $\alpha_c$. Typical aperture corrections are around $10$\,per\,cent.

\subsection{Spectral Fitting}\label{sec:models}

In \autoref{sec:fluxes} we fit the spectra of the regions, shown in \autoref{fig:finder},  derived from aperture photometry. We use a four component fit to describe the spectra defined as:
\begin{equation}\label{eqn:spectrum}
    S_\mathrm{tot} = S_\mathrm{ff} + S_\mathrm{td} + S_\mathrm{sd} + S_\mathrm{CMB},
\end{equation}
where $S_\mathrm{ff}$ describes the optically-thin free-free continuum emission from ionised gas, the thermal dust spectrum is $S_\mathrm{td}$,  $S_\mathrm{sd}$ is the spinning dust spectrum, and $S_\mathrm{CMB}$ is the integrated flux density of the CMB anisotropies. We do not include any synchrotron component in the model as previous analyses of the $\lambda$-Orionis PDR found no evidence for synchrotron emission at $1^\circ$ scales \citep{PIP_XV,CepedaArroita2021}, and the strong correlation between the optical VTSS H$\alpha$ data with GALFACTS and GB6 radio data in \autoref{fig:grid} indicate that the dominant radio emission is due just to optically thin free-free emission. We also tried directly subtracting the CMB from each frequency band using the CMB solutions from \textit{Planck} \citep{Planck2018_I} instead of fitting for the CMB flux density, but found it made little difference to the final result since the CMB is such a small component at all frequencies except in the ACT 98\,GHz band. 

We model the spectrum for optically thin free-free emission using the model given in \citet{Draine_book},
\begin{equation}
    T_\mathrm{ff} = 5.468 \times 10^{-2} \left(\frac{T_e}{\mathrm{K}}\right)^{-1.5} \left(\frac{\nu}{\mathrm{GHz}}\right)^{-2}  \left(\frac{\mathrm{EM}}{\mathrm{pc~cm}^{-6}}\right) g(T_4, \nu_\mathrm{GHz}),
\end{equation}
where $T_e$ is the electron temperature and $T_4 = T_e / 10^4$, $\mathrm{EM}$ is the emission measure, $g$ is an approximation for the Gaunt factor given by
\begin{equation}
    g(T_4, \nu_\mathrm{GHz}) = \ln\left(e^X + e^1\right),
\end{equation}
where $X = 5.960 - \sqrt{3}/\pi \ln\left(\nu_\mathrm{GHz} T_4^{-1.5}\right)$. We fit only for EM, as EM and $T_e$ are degenerate, therefore we fixed the electron temperature at a typical value for the ISM of $T_e = 8000$\,K \citep{Paladini2003}.

We fit the thermal dust emission spectrum with a modified black-body spectrum defined as
\begin{equation}\label{eqn:thermal-dust}
 S_\mathrm{td} = \tau_{353} \left(\frac{\nu}{353\mathrm{GHz}}\right)^{\beta} B_\nu(T_\mathrm{d}) \Omega,
\end{equation}
where we fit for the dust optical depth at 353\,GHz ($\tau_{353}$), the dust spectral index ($\beta$), and the effective dust temperature ($T_\mathrm{d}$). The other terms are $B_\nu$, the Planck function at temperature $T_\mathrm{d}$, and $\Omega$--the solid angle of the aperture. For the dust emissivity index $\beta$ we used a Gaussian prior of $N(1.6,0.3)$, and for the dust temperature we used $N(17,10)$, which were informed by typical all-sky values from \textit{Planck} \cite{Planck2013_XI}.
 
To model the spinning dust emission we use a warm neutral medium (WNM) template spectrum generated using the \texttt{SpDust} \citep{Ali-Hamoud2009} code using the WNM parameters defined in \citet{Draine1998a}. The template spectrum is normalised relative to the brightness at 30\,GHz. The fitted function for the spinning dust emission in then
\begin{equation}\label{eqn:spinning_dust}
S_\mathrm{sd} = A_\mathrm{sd} \frac{f(\nu \nu_p/\nu_\mathrm{sd} )}{f(\nu_r \nu_p/\nu_\mathrm{sd})} ,
\end{equation}
where we fit for the spectral flux density at 30\,GHz ($A_\mathrm{sd}$), and the spectrum peak frequency ($\nu_\mathrm{sd}$). The term $\nu_p$ is the numerically derived peak of the spinning dust spectrum, and $\nu_r$ is the reference frequency of 30\,GHz. For this analysis we have used a Gaussian prior on $\nu_\mathrm{sd}$ of $N(30\,\mathrm{GHz},10\,\mathrm{GHz})$, to help constrain the spinning dust peak to make up for the lack of data around 15 and 60\,GHz. The detailed choice of the spinning dust template does not significantly impact the results, and similar fits are obtained for cold neutral medium models (CNM). We do not use a fixed log-normal distribution, as has been used in other analyses \citep[e.g.,][]{CepedaArroita2021}, as we do not have sufficient sampling in frequency space to be able to constrain the width of the log-normal, but by using the \texttt{SpDust} WNM model the width of the spectrum is already fixed to a physically motivated value.
 
The final model we fit for is the integrated flux density from the cosmic microwave background anisotropies. We modelled these as 
\begin{equation}
	S_\mathrm{CMB} = A_\mathrm{CMB} B_\nu(T_\mathrm{CMB}) \Omega \times 10^6,
\end{equation}
where $A_\mathrm{CMB}$ is the mean CMB anisotropies amplitude in $\mu$K, $B_\nu$ is the Planck function at temperature $T_\mathrm{CMB}$, and $\Omega$ is the solid angle of the aperture. 

In summary, we performed an MCMC fit to the flux densities measured in the apertures defined by Regions \textit{Full}, A, B, C, and D---given in \autoref{tab:flux_data}. Fits were performed between 1 and 3000\,GHz resulting in 11\, data points. We fitted for seven free parameters: $\tau_{353}$, $T_\mathrm{d}$, $\beta$, $\mathrm{EM}$, $A_\mathrm{sd}$, $\nu_\mathrm{sd}$, and $A_\mathrm{CMB}$---given in \autoref{tab:modelfit}. We do not account for the correlated noise between the COMAP bands, however this should not have significant impact on the overall uncertainties as discussed in \autoref{sec:photometry}. 

For the MCMC analysis we use the affine invariant Markov chain Monte Carlo ensemble sampler (\texttt{emcee}) package \citep{Foreman-Mackey2013}. We used a total of 100 walkers, with chain lengths of 3000, and a burn-in of 1000. Chains were initialised using least-squares fits to the spectrum. We checked that all chains were converged using the Rubin-Gelman statistic. If after 3000 steps a chain was not converged, we reinitialised the chain using the average value of the converged chains. This was not required for most regions, except Region D, which required four reinitalisations before all chains were converged (giving an effective chain length of 12000).

\subsection{Integrated Flux Density}\label{sec:fluxes}

\begin{table*} 
    \caption{Flux densities for each region shown in \autoref{fig:finder} and described in \autoref{tab:apertures}. The flux densities are measured using aperture photometry and the uncertainties are calculated using bootstrapping and include calibration uncertainties.} \label{tab:flux_data}
\begin{tabular}{cccccccc}
\hline
Survey & Frequency  & \textit{Full} Region & Region A & Region B & Region \textit{Planck} & Region C & Region D   \\ 
       &  [GHz]  &  [Jy] & [Jy] & [Jy] & [Jy] & [Jy]  & [Jy] \\ 
\hline
GALFACTS & 1.447 & 0.53 $\pm$ 0.39 & 0.25 $\pm$ 0.32 & 0.24 $\pm$ 0.25 & 3.2 $\pm$ 1.3 & 0.54 $\pm$ 0.11 & 0.45 $\pm$ 0.10 \\
COMAP & 27 & 2.84 $\pm$ 0.25 & 1.41 $\pm$ 0.2 & 1.42 $\pm$ 0.2 & 7.4 $\pm$ 1.3 & 0.62 $\pm$ 0.32 & 0.56 $\pm$ 0.15 \\
COMAP & 29 & 2.93 $\pm$ 0.23 & 1.34 $\pm$ 0.22 & 1.38 $\pm$ 0.22 & 7.1 $\pm$ 1.2 & 0.48 $\pm$ 0.29 & 0.66 $\pm$ 0.13 \\
COMAP & 31 & 3.19 $\pm$ 0.26 & 1.5 $\pm$ 0.23 & 1.45 $\pm$ 0.22 & 10.3 $\pm$ 1.3 & 0.4 $\pm$ 0.44 & 0.73 $\pm$ 0.21 \\
COMAP & 33 & 2.89 $\pm$ 0.28 & 1.35 $\pm$ 0.22 & 1.16 $\pm$ 0.2 & 7.8 $\pm$ 1.5 & 0.27 $\pm$ 0.38 & 0.65 $\pm$ 0.2 \\
ACT & 98 & 3.46 $\pm$ 0.22 & 2.06 $\pm$ 0.15 & 0.83 $\pm$ 0.23 & 6.75 $\pm$ 0.95 & 1.07 $\pm$ 0.58 & 0.47 $\pm$ 0.18 \\
\textit{Planck} & 217 & 57.5 $\pm$ 8.5 & 37.1 $\pm$ 7.9 & 13.4 $\pm$ 5.9 & 178 $\pm$ 13 & 23 $\pm$ 11 & 5.6 $\pm$ 1.4 \\
\textit{Planck} & 353 & 211 $\pm$ 40 & 137 $\pm$ 37 & 54.0 $\pm$ 23.0 & 678 $\pm$ 50 & 79 $\pm$ 36 & 22.2 $\pm$ 7.1 \\
\textit{Planck} & 545 & 710 $\pm$ 110 & 450 $\pm$ 120 & 199 $\pm$ 65 & 2310 $\pm$ 170 & 250 $\pm$ 110 & 79 $\pm$ 31 \\
\textit{Planck} & 857 & 2060 $\pm$ 330 & 1340 $\pm$ 370 & 590 $\pm$ 220 & 7020 $\pm$ 500 & 750 $\pm$ 410 & 310 $\pm$ 100 \\
IRAS $100\,\mu$m & 3000 & 2200 $\pm$ 220 & 1340 $\pm$ 220 & 830 $\pm$ 180 & 6540 $\pm$ 750 & 900 $\pm$ 410 & 500 $\pm$ 150 \\
IRAS $60\,\mu$m& 5000 & 833 $\pm$ 44 & 464 $\pm$ 72 & 395 $\pm$ 70 & 2650 $\pm$ 150 & 380 $\pm$ 130 & 215 $\pm$ 49 \\
IRAS $25\,\mu$m& 12000 & 225 $\pm$ 19 & 108 $\pm$ 22 & 112 $\pm$ 15 & 676 $\pm$ 37 & 77 $\pm$ 36 & 51 $\pm$ 14 \\
WISE $22\,\mu$m& 13600 & 194 $\pm$ 17 & 93 $\pm$ 19 & 98 $\pm$ 14 & 626 $\pm$ 75 & 78 $\pm$ 38 & 36 $\pm$ 15\\
IRAS $12\,\mu$m& 25000 & 144 $\pm$ 14 & 64 $\pm$ 15 & 71 $\pm$ 13 & 431 $\pm$ 29 & 76 $\pm$ 34 & 35.7 $\pm$ 8.6 \\
WISE $12\,\mu$m& 25000 & 183 $\pm$ 28 & 70 $\pm$ 22 & 93 $\pm$ 18 & 762 $\pm$ 75 & 56 $\pm$ 25 & 42 $\pm$ 14 \\
WISE $4.6\,\mu$m& 64900 & 3.7 $\pm$ 0.3 & 1.6 $\pm$ 0.4 & 1.6 $\pm$ 0.3 & 10.3 $\pm$ 0.5 & 1.5 $\pm$ 0.4 & 0.8 $\pm$ 0.2 \\
WISE $3.4\,\mu$m& 88200 & 8.7 $\pm$ 0.9 & 4.4 $\pm$ 0.7 & 3.6 $\pm$ 0.7 & 29.7 $\pm$ 2.3 & 3.4 $\pm$ 1.0 & 3.3 $\pm$ 0.6 \\

\hline
\end{tabular}

\end{table*}

\begin{figure*}
    \centering
    \includegraphics[width=0.98\textwidth]{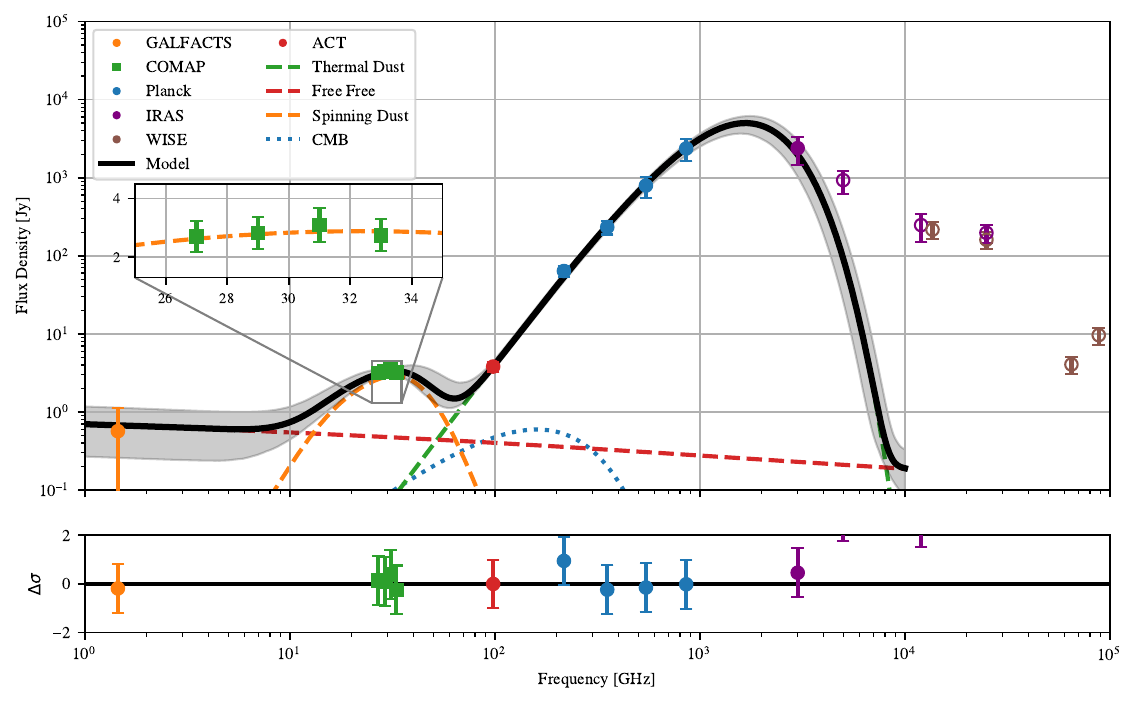}
    \caption{Best-fitting spectrum to Region \textit{Full} in \autoref{fig:finder}. The \textit{black} line indicates the integrated spectrum, while each component is shown by the \textit{dashed} lines. The amplitude of the CMB ($A_\mathrm{CMB}$) in this fit is negative, so we show the absolute value of the CMB fit using a \textit{dotted} line. The $1\sigma$ uncertainties in the model are shown by the grey highlighted region. Bands at wavelengths below 100\,$\mu$m were excluded from the fit. The inset axis shows the residual spinning dust flux density in the COMAP bands with the best-fitting spinning dust model.}
    \label{fig:aperture_spectrum}
\end{figure*}

\begin{table*}
\caption{Best-fitting parameters for each region. Uncertainties are taken from the standard deviation of the posterior distributions---such as those in \autoref{fig:corner-plot}. The spectral index of the thermal dust has Gaussian prior of $N(1.6,0.3)$ based on average all-sky values from \textit{Planck} \citep{Planck2013_XI}. The peak frequency of the spinning dust used a Gaussian prior of $N(30\,\mathrm{GHz},10\,\mathrm{GHz})$. }\label{tab:modelfit}
\centering 
\begin{tabular}{cccccccc}
\hline
Region & $\tau_{353}$  & $T_\mathrm{d}$ &  $\beta$ & EM & $A_\mathrm{sd}$ & $\nu_\mathrm{sd}$ & $A_\mathrm{CMB}$ \\ 
       &       $\times 10^{-6}$ &  [K]   & &   [cm$^{-6}$\,pc] & [Jy] & [GHz] & [$\mu$K]    \\ 
\hline
\textit{Full} & $67.0 \pm 19.0$ & $17.8 \pm 2.3$ & $1.54 \pm 0.20$ & $71 \pm 47$ & $2.85 \pm 0.43$ & $32.3 \pm 5.9$ & $-17 \pm 35$ \\
\textit{Planck} & $37.8 \pm 9.3$ & $18.0 \pm 2.3$ & $1.49 \pm 0.20$ & $57 \pm 29$ & $7.30 \pm 1.40$ & $33.2 \pm 6.4$ & $-48 \pm 23$ \\
A & $75.0 \pm 26.0$ & $17.5 \pm 2.5$ & $1.57 \pm 0.21$ & $62 \pm 44$ & $1.32 \pm 0.26$ & $31.6 \pm 6.7$ & $-13 \pm 38$ \\
B & $23.0 \pm 12.0$ & $19.2 \pm 3.5$ & $1.58 \pm 0.27$ & $86 \pm 50$ & $1.19 \pm 0.25$ & $29.0 \pm 6.5$ & $-10 \pm 28$ \\
C & $41.0 \pm 28.0$ & $17.9 \pm 4.6$ & $1.54 \pm 0.27$ & $99 \pm 37$ & $0.27 \pm 0.20$ & $29.6 \pm 9.3$ & $-19 \pm 39$ \\
D & $9.9 \pm 5.1$ & $20.6 \pm 4.9$ & $1.51 \pm 0.27$ & $87 \pm 34$ & $0.41 \pm 0.17$ & $32.7 \pm 8.3$ & $-11 \pm 18$\\\hline
\end{tabular}
\end{table*}

\begin{figure*}
    \centering
    \includegraphics[width=0.98\textwidth]{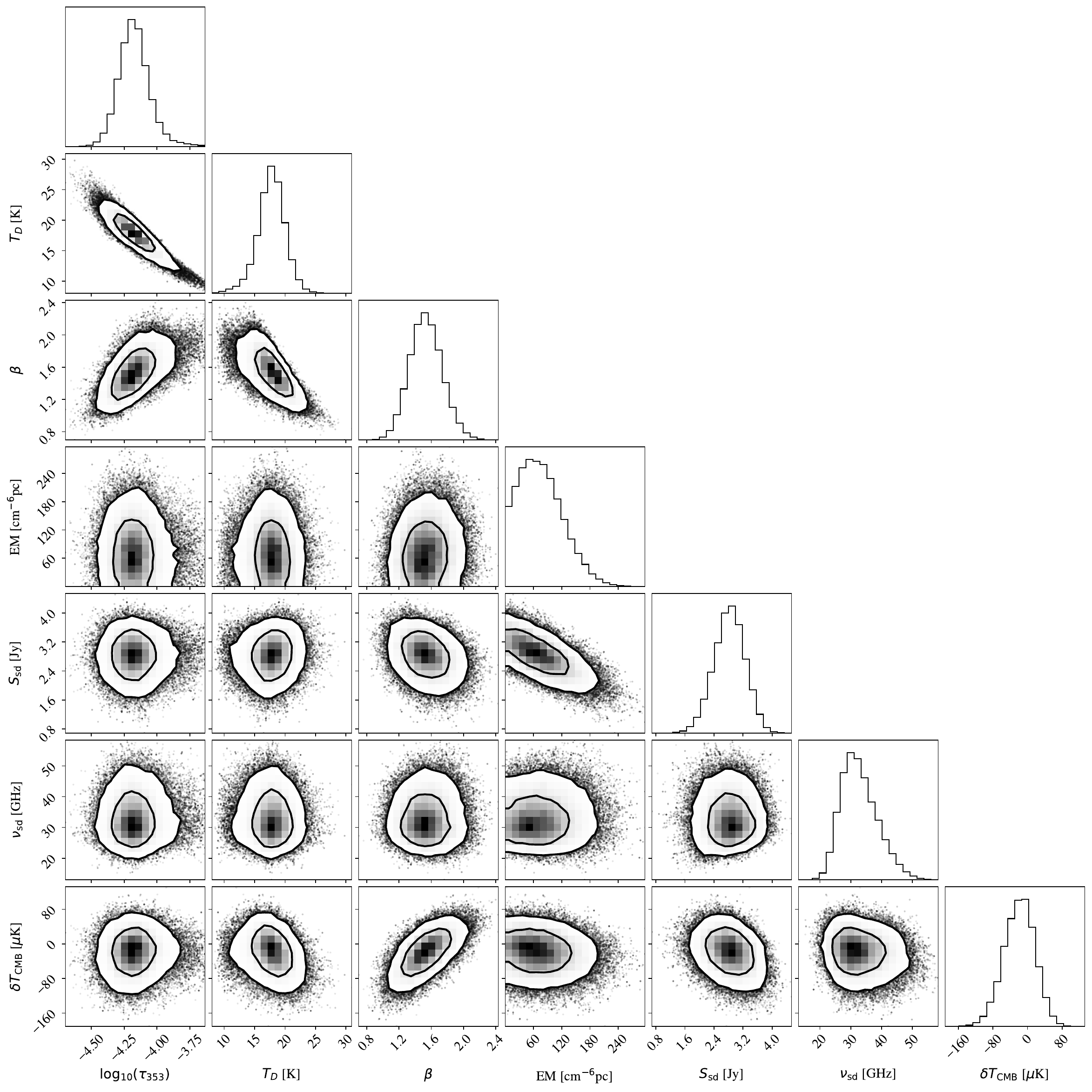}
    \caption{Posterior distributions for each fitted parameter for the spectrum of Region \textit{Full} shown in \autoref{fig:aperture_spectrum}. Contours are shown at $1$ and $2\sigma$.}
    \label{fig:corner-plot}
\end{figure*}

\autoref{fig:aperture_spectrum} shows the best-fitting model of \autoref{eqn:spectrum} to the photometry data measured within Region \textit{Full} shown in \autoref{fig:finder}, and \autoref{fig:corner-plot} shows the posterior distributions. \autoref{tab:flux_data} gives the measured flux densities at each frequency. \autoref{fig:aperture_spectrum} also shows the model residuals normalised by the uncertainties in each flux density. The dust emitting at frequencies above the IRAS\,100\,$\mu$m band are much more susceptible to stochastic heating that is not well described by a single-temperature modified black-body spectrum. These bands are therefore excluded from the fit (shown as hollow markers). The model is most uncertain between 50--100\,GHz due to a lack of data at the required angular resolution at these frequencies.

The best-fitting models for each of the three regions are given in \autoref{tab:modelfit}. The uncertainties in the table are the standard deviations of each parameter's posterior distribution. Spinning dust is detected at a high significance of $7\sigma$ in the \textit{Full} region. This corresponds to a total fraction of AME at 30\,GHz of $86.0 \pm 0.08$\,per\,cent.

The uncertainties in \autoref{fig:aperture_spectrum} are calculated by measuring the distribution of flux densities when moving the aperture within the field, as discussed in \autoref{sec:photometry}. The $\chi^2$ of the residuals for the fit are $\chi^2 = 1.54$ with a reduced $\chi^2_r = 0.39$ (4 degrees-of-freedom). The small $\chi^2_r$ is because we do not account for the correlations in the uncertainties. For COMAP, the noise between frequency bands is highly correlated ($\approx 90$\,\%), meaning we should really only consider COMAP to be a single data point in \autoref{fig:aperture_spectrum}. If we do this the degrees-of-freedom becomes approximately unity, and the reduced $\chi^2_r$ becomes $1.54$. We also do not model how the background emission is correlated between bands in the sub-mm and infrared bands, and in \autoref{fig:aperture_spectrum} we show these as entirely independent uncertainties. This has the effect of increasing the effective uncertainties of the model fit, and to decrease the $\chi^2$ of the fit. 

We used the Bayesian information criterion (BIC) to compare models with and without a  spinning dust component. With spinning dust the model $\mathrm{BIC}_\mathrm{sd} = 111$, while without spinning dust $\mathrm{BIC}_\mathrm{ff} = 145$. For hierarchical models, the lower BIC value is favoured if the difference is greater than 5. In the case the difference is $\Delta \mathrm{BIC} = 34$ suggesting the data strongly favour the inclusion of an spinning dust component \citep{Hilbe2017}.

\subsubsection{Comparison with Planck}

The Region \textit{Planck} aperture was chosen to encompass all of the emission seen within the COMAP data so that it can be directly compared with the flux density measured in the \textit{Planck} survey of Galactic spinning dust emission regions \citep{PIP_XV}. The \textit{Planck} analysis used a $1^\circ$ radius aperture at $1^\circ$ resolution, and reported a flux density of $7.6 \pm 1.0$\,Jy at 28.4\,GHz for the B30 region, while we find (scaling the value from \autoref{tab:modelfit}) a flux density of $6.8 \pm 1.4$\,Jy at 28.4\,GHz (we do not account for colour corrections in this comparison as they are negligible). This implies that most of the emission seen in the lower resolution \textit{Planck} 28.4\,GHz data is due to spinning dust emission within the B30 cloud that is seen by COMAP. This is approximately twice the flux density measured within the tighter aperture that just captures the flux from the central cloud, which has a flux density of $S_\mathrm{Full} = 2.68 \pm 0.37$\,Jy at 28.4\,GHz (again, scaling the value from \autoref{tab:modelfit}). This implies that approximately half of the spinning dust emission within B30, when measured at $1^\circ$ scales, is due to a diffuse component outside of the Region \textit{Full} aperture shown in \autoref{fig:finder}.

One issue when comparing with the $1^\circ$ resolution \textit{Planck} analysis is that its $1^\circ$ radius aperture is much larger than is possible for the COMAP data and so would have integrated a larger fraction of the surrounding free-free emission alongside the spinning dust component. In \citet{PIP_XV} the average emission measure within the measured aperture was $\mathrm{EM} = 8 \pm 6$\,pc\,cm$^{-6}$. We can use this, along with the reported spinning dust flux density of $7.6 \pm 1.0$\,Jy, to estimate the fraction of spinning dust emission to the total emission at 28.4\,GHz. We find that the \textit{Planck} spinning dust flux density ratio is $80 \pm 14$\,per\,cent, while we found the ratio with the \textit{Planck} region aperture to be much higher at $95 \pm 3$\,per\,cent. This difference is driven by the fact we are using a slightly smaller aperture that integrates less of the $\lambda$-Orionis H$\textsc{ii}$ region, and shows that the \textit{Planck} analysis underestimated the fraction of the emission from B30 that is due to spinning dust emission. 

\subsubsection{Comparison of GALFACTS with VTSS}

In \autoref{tab:flux_data} we show that the 1.447\,GHz GALFACTS data within the \textit{Full}, A and B aperture regions is consistent with noise. Looking at the GALFACTS data in \autoref{fig:grid} it is clear that there is a free-free background that forms a gradient across the region.

Within the measured apertures free-free emission does not make a significant contribution. This is also clear if one compares the GALFACTS and COMAP data in \autoref{fig:grid}, there is no clear correlation between the two data sets. In Region \textit{Full} the spinning dust emission detected at 30\,GHz is approximately $7\sigma$ above the radio free-free emission predicted by the GALFACTS data---meaning the GALFACTS data would need to be almost an order-of-magnitude brighter to match the flux density we see at 30\,GHz. 

We can cross-check the calibration of the GALFACTS data by comparing it to the VTSS H$\alpha$ data. Both H$\alpha$ and radio free-free emission are generated by the same ionised gas and a direct relationship between the H$\alpha$ intensity and free-free brightness is expected \citep{Draine_book}. Although we do not correct for dust absorption, significant H$\alpha$ extinction is only apparent for certain regions within B30 but since we are only interested in the average relationship between the radio and H$\alpha$ emission over the whole region the impact of dust absorption should be minimal. We measured the gradient between between the VTSS and GALFACTS data within the common pixels shown in \autoref{fig:grid} and found $I(\mathrm{H}\alpha)/T_\mathrm{ff} = 5.88 \pm 0.04$\,mK/R, while the expected ratio at 1.447\,GHz (assuming an electron temperature between 5000--10000\,K) is $I(\mathrm{H}\alpha)/T_\mathrm{ff} = 3.0$--$6.6$\,mK/R \citep[e.g.,][]{Dickinson2003}. Therefore, the free-free level is consistent between the radio and optical data.

\begin{table*}
    \centering
        \caption{Derived values from the photometry and model fits of the different regions shown in \autoref{fig:finder}. The second and third columns are the spinning dust emissivity relative to the fitted spinning dust brightness at 30\,GHz, the fourth column is the average relative interstellar radiation field (ISRF) in the region, and columns five and six are the ratios of the flux densities between the WISE 12 and 4.6\,$\mu$m bands and the WISE 22\,$\mu$m band.}
\begin{tabular}{ccccccc}
\hline
Region & T$_{30\mathrm{GHz}}$/$\tau_{353}$  & T$_{30\mathrm{GHz}}$/IRAS$_{100\mu\mathrm{m}}$ & $G$ & $\alpha^{22\mu\mathrm{m}}_{4.6\mu\mathrm{m}}$ & $f^\mathrm{PAH}_{12\mu\mathrm{m}}$ & $f^\mathrm{PAH}_{3.4\mu\mathrm{m}}$\\ 
       &  [K]                                 &  [$\mu$K/(MJy/sr)] & & &  &  \\ 
\hline
\textit{Full} & $22.1 \pm 5.0$ & $53.0 \pm 10.0$ & $1.1 \pm 0.8$ & $-2.55 \pm 0.08$ & $0.31 \pm 0.07$ & $0.24 \pm 0.06$ \\
\textit{Planck} & $16.9 \pm 3.6$ & $40.7 \pm 9.5$ & $1.2 \pm 0.8$ & $-2.59 \pm 0.09$ & $0.20 \pm 0.04$ & $0.19 \pm 0.05$ \\
A & $15.6 \pm 5.0$ & $41.0 \pm 11.0$ & $1.0 \pm 0.8$ & $-2.58 \pm 0.22$ & $0.37 \pm 0.20$ & $0.19 \pm 0.11$ \\
B & $43.0 \pm 21.0$ & $59.0 \pm 18.0$ & $1.7 \pm 1.7$ & $-2.65 \pm 0.16$ & $0.29 \pm 0.10$ & $0.25 \pm 0.12$ \\
C & $5.7 \pm 5.4$ & $11.4 \pm 9.6$ & $1.1 \pm 1.6$ & $-2.55 \pm 0.46$ & $0.29 \pm 0.23$ & $0.26 \pm 0.22$ \\
D & $39.0 \pm 23.0$ & $29.0 \pm 15.0$ & $2.5 \pm 3.2$ & $-2.39 \pm 0.41$ & $0.24 \pm 0.18$ & $0.13 \pm 0.12$ \\\hline
\end{tabular}
    \label{tab:derived_products}
\end{table*}

\subsubsection{Differences Between Region A and B} 

\begin{figure}
    \centering
    \includegraphics[width=\columnwidth]{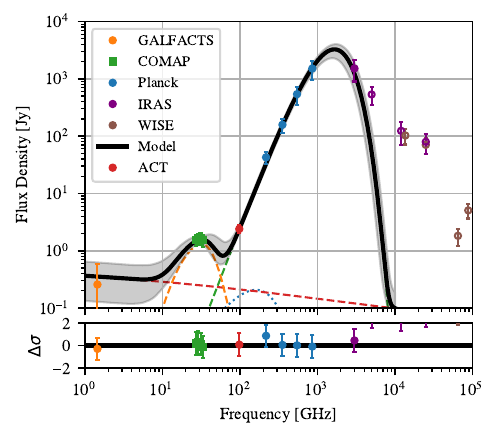}
    \includegraphics[width=\columnwidth]{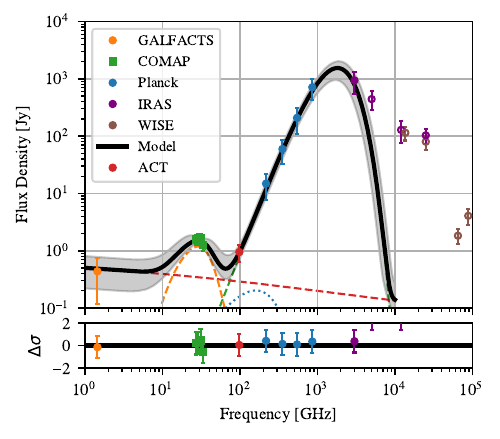}
    \caption{Best-fitting spectra to Region A (\textit{top}) and Region B (\textit{bottom}) in \autoref{fig:finder}. The \textit{black} line indicates the integrated spectrum, while each component is shown by the \textit{dashed} lines. The $1\sigma$ uncertainties in the model are shown by the grey highlighted region. Bands at wavelengths below 100\,$\mu$m were excluded from the fit. }
    \label{fig:RegionAandBSpectrum}
\end{figure}

In \autoref{fig:RegionAandBSpectrum} we show the spectra for both Region A and Region B, which are the two bright spinning dust emitting hot spots in the COMAP data. Comparing these to \autoref{fig:aperture_spectrum} it is clear that all three spectra are very similar, with the same spinning dust emission peak brightness, and free-free radio background. The largest difference in both spectra comes from the measurements of the thermal dust emission---in Region B this is more than a factor of two lower than in Region A. However, as illustrated by the figures and \autoref{tab:flux_data}, the far-infrared (FIR) data at 100 and 60\,$\mu$m and the mid-infrared (MIR) data at $\lambda < 25$\,$\mu$m are consistent within uncertainties. This is also clear in \autoref{fig:grid} where the relative brightness of Region A is much brighter than Region B, but in the mid-infrared bands this difference between Region A and B becomes less apparent---similar to what is seen in the COMAP data. 

How can we explain why the flux density of the spinning dust emission and MIR data are equal in Region A and B, while the sub-mm thermal dust flux densities vary by a factor of two? It suggests that the total dust column, as traced by the thermal dust emission, is not an accurate tracer of the spinning dust carriers at parsec scales. We know that Region A has a much higher dust column density than Region B, both from the peak in the \textit{Planck} bands shown in \autoref{fig:grid}, which is a known cold clump \citep{planck2015_XXVIII}, and from other observations that have identified cold cores within the region \citep{Kauffmann2008}. Therefore the spinning dust carriers must be part of the lower density, warm dust phase within B30, which is better traced by the FIR and MIR data. 

In \autoref{tab:derived_products} we provide several statistics derived from the photometry and the model fits. The first two columns of \autoref{tab:derived_products} show the spinning dust emissivity derived for both the brightness of spinning dust emission at 30\,GHz over $\tau_{353}$ and the IRAS 100\,$\mu$m flux density. It is useful to compare the 30\,GHz flux densities with the dust optical depth at 353\,GHz ($\tau_{353}$), which is thought to be one of the best direct tracers of the total dust column along a line-of-sight \citep{Planck2016_XXIX}. 

Typical values for the emissivity of spinning dust relative to $\tau_{353}$ ranges from $5-25$\,K at 28.4\,GHz, with higher values in denser molecular cloud regions such as Perseus or $\rho$-Ophiuchus \citep{Planck2011_XX} and lower values in lower density high-latitude cirrus dust clouds \citep{Harper2022}. We find that at 28.4\,GHz the spinning dust emissivity per unit $\tau_{353}$ is statistically consistent between all regions. The \textit{Full} region has an emissivity of $20.1 \pm 4.8$\,K, while Region A and B have emissivities of $14.9 \pm 5.0$\,K and $44.0 \pm 21.0$\,K, respectively (these are scaled from the 30\,GHz values given in \autoref{tab:derived_products}). 


When deriving the spinning dust emissivity relative to IRAS 100\,$\mu$m data, we find that the two regions are more consistent ($41 \pm 11$ and $59 \pm 18$\,$\mu$K/(MJy/sr), respectively). However, both regions have an emissivity that is higher than any other observed PDR associated with spinning dust emission measured to date, which typically have values between 10 and 25\,$\mu$K/(MJy/sr) \citep[e.g.,][]{Casassus2006,Dickinson2007,Planck2011_XX}.

In \autoref{tab:modelfit} we can see that the average dust temperature within Region A is lower than in Region B, which, assuming both clouds have similar grain populations, implies that Region B has a more intense total interstellar radiation field (ISRF). We give estimates of the ISRF ($G_0$) in \autoref{tab:derived_products} using
\begin{equation}
    G_0 = \left(\frac{T_d}{17.5\mathrm{K}}\right)^{4 + \beta},
\end{equation}
where $\beta$ and $T_\mathrm{d}$ are given in \autoref{tab:modelfit}. Region A has a lower average $G_0$ than Region B, implying there is more shielding in this region that could lead to grain coagulation and depletion of the smallest grains that are carriers of spinning dust emission. 

Finally, in \autoref{tab:derived_products} we estimate the PAH emission fraction in each region. We do this by fitting for a power-law between the 4.6 and 22\,$\mu$m bands, as both of these bands are free of bright PAH features and should be dominated by the continuum background (however these bands may contain ionic PAH features that we are not accounting for here). This gives us a function for the continuum background of $S_\mathrm{c}(\lambda) = S_{22\mu\mathrm{m}} \left(\frac{22\,\mu\mathrm{m}}{4.6\,\mu\mathrm{m}}\right)^\alpha$, where the $\alpha$ values are given in \autoref{tab:derived_products}. The fraction of PAHs is then defined as 
\begin{equation}
    f_\lambda^\mathrm{PAH} = 1 - \frac{S_\lambda}{S_\mathrm{c}(\lambda)},
\end{equation}
where $S_\lambda$ is the flux measured in either the 3.6 or 12\,$\mu$m WISE bands. We find that both regions do contain a significant PAH fraction, between 20 and 40\,per\,cent depending on the region and WISE band used. However, we do not find that the PAH fraction is significantly different between both regions. 

\subsubsection{Region C and Region D} 

\begin{figure}
    \centering
    \includegraphics[width=\columnwidth]{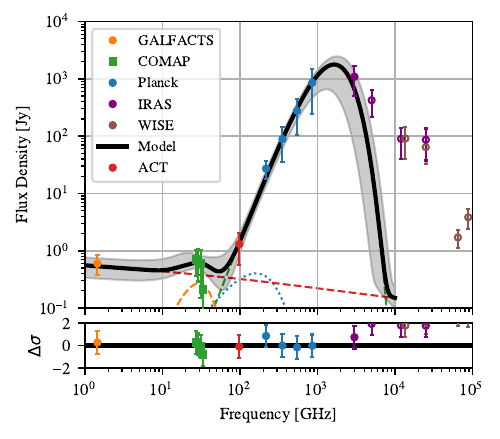}
    \includegraphics[width=\columnwidth]{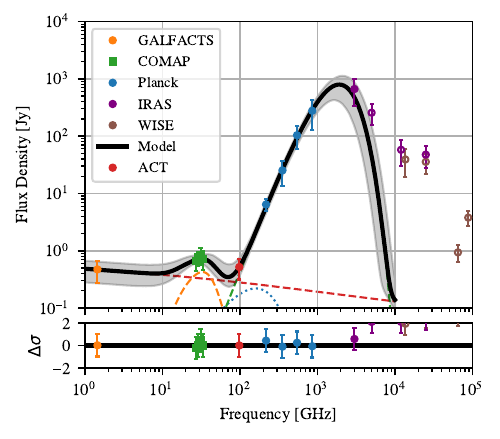}
    \caption{Best-fitting spectra to Region C (\textit{top}) and Region D (\textit{bottom}) in \autoref{fig:finder}. The \textit{black} line indicates the integrated spectrum, while each component is shown by the \textit{dashed} lines. The $1\sigma$ uncertainties in the model are shown by the grey highlighted region. Bands at wavelengths below 100\,$\mu$m were excluded from the fit.}
    \label{fig:RegionCandDSpectrum}
\end{figure}

In \autoref{fig:RegionCandDSpectrum} we show the spectra for both Region C and Region D, two additional sub-regions of B30. Region C is known to have significant low-mass star formation and multiple young-stellar objects (YSOs), proto-brown dwarfs, and prestellar cores have been identified \citep{Huelamo2017,Barrado2018}. The region protrudes inward towards the $\lambda$-Orionis star, and as such has the most intense ISRF (as seen by $G_0$ in \autoref{tab:derived_products} and $T_\mathrm{d}$ in \autoref{tab:modelfit}), and the highest background contamination due to radio free-free emission---Region C can be seen as an absorption feature within the VTSS data shown in \autoref{fig:grid}. However, in terms of the dust column, Region C is similar to both regions A and B (see $\tau_{353}$ in \autoref{tab:modelfit}) yet the spinning dust flux density in this region is significantly lower and is consistent with zero. In fact, the BIC for the spectrum including spinning dust emission shown in \autoref{fig:RegionCandDSpectrum} is $\mathrm{BIC}=17$ and without spinning dust emission it is $\mathrm{BIC} = 12$, which strongly suggests that the no spinning dust model is preferred. 

Region D is a smaller cloud separate from the main B30 complex. It is coincident with a faint (0.072\,Jy at 1.4\,GHz) radio source \citep{Healey2009}, contains some low-mass star formation, and the lowest thermal dust optical depth of all of the regions. Although it is fainter than the main cloud in all bands, this region has the same spinning dust emissivity per $\tau_{353}$ (\autoref{tab:derived_products}) as Region B, but the same spinning dust emissivity per $I(100\mu\mathrm{m})$ as Region A. The implication being that the spinning dust carriers are likely traced by the warmer, FIR and MIR dust population in this region rather than the total thermal dust population (as traced by $\tau_{353}$).

\section{Template Fitting}\label{sec:template-fitting}

As well as looking at the integrated flux density of B30, we use a template fitting or correlation analysis to decompose the COMAP bands into the constituent emission components---free-free and dust emission. Template fitting methods have been used extensively for studying diffuse Galactic emission \citep[e.g.][]{Kogut1996, deOliveira-Costa1997, Davies2006, Peel2012, Harper2022}. The template fitting method assumes that the data can be decomposed into a linear combination of spatial templates as 
\begin{equation}\label{eqn:template-model}
    \mathbf{d} = \mathbf{A} \mathbf{s} + \mathbf{n},
\end{equation}
where $\mathbf{d}$ is the data, $\mathbf{A}$ is the template matrix, $\mathbf{s}$ is the vector of template coefficients, and $\mathbf{n}$ is the noise vector. The maximum likelihood estimate of the template coefficients are then solved for by 
\begin{equation}
    \mathbf{s} = \left(\mathbf{A}^T \mathbf{N}^{-1} \mathbf{A}\right)^{-1} \mathbf{A}^T \mathbf{N}^{-1} \mathbf{d},
\end{equation}
where $\mathbf{N}$ is the noise covariance matrix. The noise covariance matrix is defined as $\mathbf{N} = \langle \mathbf{n} \mathbf{n}^T \rangle$. Nominally the uncertainties in $\mathbf{s}$ can be found by taking the trace of $\left(\mathbf{A}^T \mathbf{N}^{-1} \mathbf{A}\right)$, however, for this analysis we use the same technique described in \citet{Harper2022}, where the uncertainties are bootstrapped by resampling the template region. Uncertainties for the template fitting results include only the statistical uncertainties due to the noise in each map, and the calibration uncertainty of each map. The template fitting uncertainties are often much smaller than the uncertainties in the aperture photometry as they do not include photometry errors due to the background emission.

For this analysis we only fit two templates to each map: a dust template and a geometric offset. We do not fit directly for the free-free emission because the level of free-free contamination at 30\,GHz is small, this is evident in \autoref{fig:grid} where the morphology of the central B30 cloud as seen at 30\,GHz has very little correlation with the free-free emission seen by GALFACTS, GB6 or the H$\alpha$ data. As the free-free is so low at 30\,GHz, we found that fitting for it directly introduces systematic biases in the derived dust template coefficients that vary depending on the dust template used. Specifically, the correlation between dust and free-free emission is positive when using the \textit{Planck} bands and negative for the IRAS bands, resulting in a change in the expected free-free brightness by up to 50\,per\,cent. Therefore to account for the free-free contribution in the COMAP bands we subtract the GALFACTS 1.447\,GHz data scaled by a typical optically thin free-free spectrum of 
\begin{equation} 
    T_\mathrm{ff} = T_\mathrm{GALFACTS} \left(\frac{1.447}{\nu_\mathrm{GHz}}\right)^{-2.1},
\end{equation}
where $T_\mathrm{ff}$ is the expected free-free contribution at frequency $\nu_\mathrm{GHz}$, and $T_\mathrm{GALFACTS}$ are the GALFACTS pixel brightnesses. 

The final step before template fitting the data was to fit a 2D Gaussian model to a source in Region C seen in the sub-mm, FIR, and MIR bands as the aperture photometry suggests it does not have a spinning dust component (as discussed in \autoref{sec:photometry}). Subtracting out the source, rather than just masking it, ensures that there is not a large negative systematic in the residual maps after removing the free-free model and best-fitting dust models from the COMAP data.

\subsection{Spectral Models}\label{sec:template-fit-models}

The natural units for the template fitting analysis are in units of K per unit dust template. However, we convert these into units of MJy/sr per MJy/sr dust template so that we can use the same dust models described in \autoref{sec:models}. 

To model the template fitted dust coefficients we used 
\begin{equation}
    \mathbf{S}_\mathrm{d} = \mathbf{S}_\mathrm{td} + \mathbf{S}_\mathrm{sd},
\end{equation}
where $\mathbf{S}_\mathrm{td}$ and $\mathbf{S}_\mathrm{sd}$ are the same as given in \autoref{eqn:thermal-dust} and \autoref{eqn:spinning_dust}, respectively. 

We use only positivity priors on the spinning dust amplitude and thermal dust optical depth. For the dust temperature we use a uniform prior of $0 < T_D < 100$\,K. For $\beta$ we use a Gaussian prior of $N(1.6,0.3)$, where the mean of the prior is $\beta = 1.6$, which \textit{Planck} found to be the all-sky average value \citep{Planck2013_XI}.

\subsection{Fitted Template Coefficients}\label{sec:template-coefficients}

\begin{figure}
    \centering
    \includegraphics[width=\columnwidth]{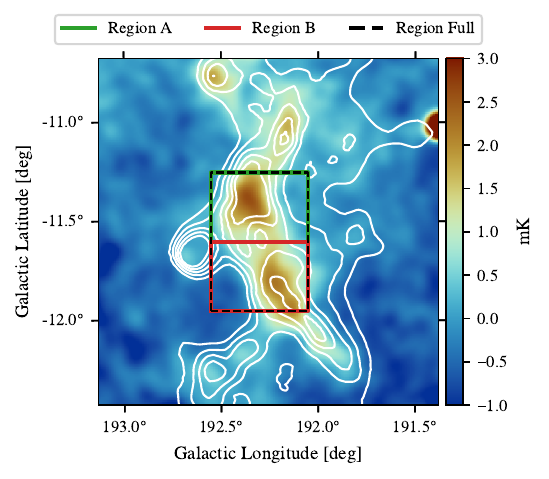}
    \caption{The COMAP 27\,GHz data with free-free emission subtracted. The contours are from the IRAS 100\,$\mu$m data at 30, 40, 50, 60, and 70\,MJy\,sr$^{-1}$. The boxed regions indicate the areas used for the template fitting analysis. The areas of both Region A and B are $0.\!\!^\circ5 \times 0.\!\!^\circ35$. }
    \label{fig:template-regions}
\end{figure}

\begin{figure*}
    \centering
    \includegraphics[width=0.98\textwidth]{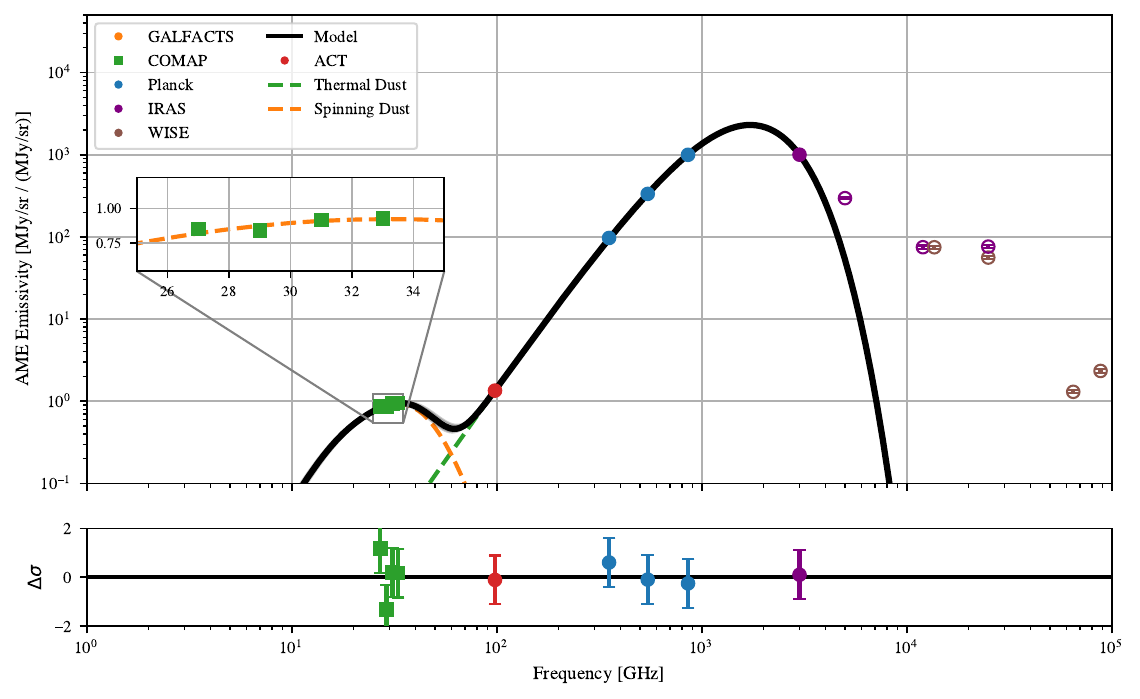}
    \caption{Best-fitting spectrum to the IRAS 100\,$\mu$m dust template coefficients for Region \textit{Full} shown in \autoref{fig:template-regions}. The \textit{black} line indicates the integrated spectrum, while each component is shown by the \textit{dashed} lines. The $1\sigma$ uncertainties in the model are shown by the grey highlighted region---in this case these are smaller than the line width of the black line. Bands at wavelengths shorter than 100\,$\mu$m were excluded from the fit. The inset axis shows the residual template coefficients in the COMAP bands after subtracting the best-fitting thermal dust model.}
    \label{fig:template_spectrum}
\end{figure*}

\begin{table*}
    \centering
    \caption{Best-fitting spinning dust and thermal dust model parameters to the dust template coefficients for the \textit{Full} region in \autoref{fig:template-regions}. The spectral index $\alpha^{27\mathrm{GHz}}_{33\mathrm{GHz}}$ is measured within the COMAP band from the fitted dust template coefficients in flux density units. The Pearson correlations at 27 and 33\,GHz are given by $r_{27\mathrm{GHz}}$ and $r_{33\mathrm{GHz}}$, respectively.}
    \label{tab:template-coefficients}
\renewcommand{\arraystretch}{1.5} 
\begin{threeparttable}
\begin{tabular}{cccccc|ccc}
\hline
Frequency & Wavelength & $S_{30\mathrm{GHz}}$ & Peak Frequency & $\beta$ & $T_d$ & $\alpha^{27\mathrm{GHz}}_{33\mathrm{GHz}}$ & $r_{27\mathrm{GHz}}$ & $r_{33\mathrm{GHz}}$ \\ 
      \.[GHz]&  & [$\mu$K/(MJy/sr)]   & [GHz]          &          & [K]   & & & \\ 
\hline
98 & 3.1mm & $22560 \pm 590$ & $33.7\pm2.5$ & $1.58\pm0.02$ & $18.1 \pm 0.2$ & $0.62 \pm 0.34$ & $0.88^{0.89}_{0.87}$ & $0.88^{0.89}_{0.87}$\\
353 & 0.8mm & $271.5 \pm 9.3$ & $35.4\pm2.6$ & $1.65\pm0.02$ & $17.3 \pm 0.2$ & $1.14 \pm 0.43$ & $0.80^{0.81}_{0.79}$ & $0.90^{0.91}_{0.89}$\\
545 & 0.6mm & $79 \pm 3$ & $35\pm3$ & $1.65\pm0.02$ & $17.3 \pm 0.2$ & $1.10 \pm 0.60$ & $0.80^{0.82}_{0.79}$ & $0.90^{0.91}_{0.89}$\\
857 & 0.3mm & $27.3 \pm 0.9$ & $35.2\pm2.5$ & $1.65\pm0.02$ & $17.4 \pm 0.2$ & $1.20 \pm 0.40$ & $0.82^{0.83}_{0.81}$ & $0.91^{0.92}_{0.90}$\\
3000 & 100$\mu$m & $32.4 \pm 0.5$ & $33.0\pm1.5$ & $1.61\pm0.03$ & $18.1 \pm 0.2$ & $0.60 \pm 0.20$ & $0.95^{0.96}_{0.94}$ & $0.94^{0.95}_{0.93}$\\
5000 & 60$\mu$m & $103.7 \pm 1.7$ & $30.2\pm1.1$ & $1.57\pm0.04$ & $18.7 \pm 0.3$ & $0.17 \pm 0.24$ & $0.96^{0.97}_{0.95}$ & $0.87^{0.88}_{0.86}$\\
12000 & 25$\mu$m & $362 \pm 9$ & $29.2\pm1.5$ & $1.55\pm0.07$ & $19.1 \pm 0.5$ & $-0.14 \pm 0.33$ & $0.93^{0.94}_{0.92}$ & $0.81^{0.82}_{0.79}$\\
13600 & 22$\mu$m & $384 \pm 8$ & $30.4\pm1.4$ & $1.57\pm0.06$ & $18.8 \pm 0.4$ & $0.19 \pm 0.27$ & $0.93^{0.94}_{0.92}$ & $0.86^{0.87}_{0.84}$\\
25000 & 12$\mu$m & $474 \pm 11$ & $30.2\pm1.7$ & $1.57\pm0.06$ & $18.9 \pm 0.4$ & $0.14 \pm 0.33$ & $0.91^{0.92}_{0.90}$ & $0.82^{0.83}_{0.81}$\\
64900 & 4.6$\mu$m & $21230 \pm 530$ & $29.8\pm1.6$ & $1.56\pm0.06$ & $18.8 \pm 0.5$ & $0.04 \pm 0.30$ & $0.93^{0.94}_{0.92}$ & $0.82^{0.84}_{0.81}$\\
88200 & 3.4$\mu$m & $11440 \pm 310$ & $29.0\pm1.5$ & $1.55\pm0.06$ & $19.1 \pm 0.5$ & $-0.20 \pm 0.30$ & $0.94^{0.95}_{0.93}$ & $0.81^{0.83}_{0.80}$\\\hline
\end{tabular}
\begin{tablenotes}
\item[a] The spinning dust emissivity per unit 353\,GHz dust opacity is $10.55 \pm 0.36$.
\end{tablenotes}
\end{threeparttable}
\end{table*}

\begin{table*}
    \centering
    \caption{Best-fitting spinning dust and thermal dust model parameters to the dust template coefficients for the Region A in \autoref{fig:template-regions}. The spectral index $\alpha^{27\mathrm{GHz}}_{33\mathrm{GHz}}$ is measured within the COMAP band from the fitted dust template coefficients in flux density units. The Pearson correlations at 27 and 33\,GHz are given by $r_{27\mathrm{GHz}}$ and $r_{33\mathrm{GHz}}$, respectively.}
    \label{tab:template-coefficients-RegionA}
\renewcommand{\arraystretch}{1.5} 
\begin{threeparttable}
\begin{tabular}{cccccc|ccc}
\hline
Frequency & Wavelength & $S_{30\mathrm{GHz}}$ & Peak Frequency & $\beta$ & $T_d$ & $\alpha^{27\mathrm{GHz}}_{33\mathrm{GHz}}$ & $r_{27\mathrm{GHz}}$ & $r_{33\mathrm{GHz}}$ \\ 
      \.[GHz]&  & [$\mu$K/(MJy/sr)]   & [GHz]          &          & [K]   & & & \\ 
\hline
98 & 3.1mm & $18410 \pm 590$ & $28.3\pm1.7$ & $1.60\pm0.04$ & $17.92 \pm 0.36$ & $-0.31 \pm 0.37$ & $0.88^{0.89}_{0.87}$ & $0.84^{0.86}_{0.82}$\\
353 & 0.8mm & $216 \pm 9.5$ & $32.6\pm3.0$ & $1.67\pm0.02$ & $16.81 \pm 0.22$ & \,$0.54 \pm 0.56$ & $0.77^{0.79}_{0.75}$ & $0.87^{0.89}_{0.85}$\\
545 & 0.6mm & $63 \pm 3$ & $32.7\pm3.0$ & $1.68\pm0.02$ & $16.82 \pm 0.2$ & \,$0.59 \pm 0.49$ & $0.77^{0.79}_{0.75}$ & $0.87^{0.89}_{0.86}$\\
857 & 0.3mm & $22.0 \pm 0.9$ & $32.5\pm2.7$ & $1.68\pm0.02$ & $16.83 \pm 0.2$ & \,$0.53 \pm 0.48$ & $0.78^{0.79}_{0.75}$ & $0.88^{0.89}_{0.87}$\\
3000 & 100$\mu$m & $30.1 \pm 0.8$ & $29.5\pm1.6$ & $1.64\pm0.04$ & $17.71 \pm 0.24$ & $-0.03 \pm 0.30$ & $0.93^{0.94}_{0.92}$ & $0.94^{0.95}_{0.93}$\\
5000 & 60$\mu$m & $95.9 \pm 2.4$ & $28.2\pm1.6$ & $1.56\pm0.07$ & $18.61 \pm 0.62$ & $-0.45 \pm 0.37$ & $0.94^{0.95}_{0.93}$ & $0.87^{0.88}_{0.86}$\\
12000 & 25$\mu$m & $422 \pm 12$ & $28.0\pm1.6$ & $1.56\pm0.11$ & $18.84 \pm 0.88$ & $-0.53 \pm 0.43$ & $0.93^{0.94}_{0.92}$ & $0.84^{0.86}_{0.82}$\\
13600 & 22$\mu$m & $459 \pm 13$ & $28.8\pm1.7$ & $1.61\pm0.09$ & $18.4 \pm 0.68$ & $-0.32 \pm 0.41$ & $0.92^{0.93}_{0.91}$ & $0.87^{0.89}_{0.86}$\\
25000 & 12$\mu$m & $622 \pm 18$ & $29.1\pm1.9$ & $1.61\pm0.09$ & $18.31 \pm 0.69$ & $-0.22 \pm 0.43$ & $0.90^{0.91}_{0.89}$ & $0.87^{0.88}_{0.85}$\\
64900 & 4.6$\mu$m & $25220 \pm 640$ & $28.6\pm1.6$ & $1.59\pm0.08$ & $18.37 \pm 0.65$ & $-0.51 \pm 0.43$ & $0.93^{0.94}_{0.92}$ & $0.87^{0.88}_{0.85}$\\
88200 & 3.4$\mu$m & $14620 \pm 330$ & $28.5\pm1.4$ & $1.58\pm0.10$ & $18.61 \pm 0.83$ & $-0.51 \pm 0.37$ & $0.95^{0.96}_{0.94}$ & $0.88^{0.89}_{0.86}$\\\hline
\end{tabular}
\begin{tablenotes}
\item[a] The spinning dust emissivity per unit 353\,GHz dust opacity is $8.05 \pm 0.35$.
\end{tablenotes}
\end{threeparttable}
\end{table*}

\begin{table*}
    \centering
    \caption{Best-fitting spinning dust and thermal dust model parameters to the dust template coefficients for the Region B in \autoref{fig:template-regions}. The spectral index $\alpha^{27\mathrm{GHz}}_{33\mathrm{GHz}}$ is measured within the COMAP band from the fitted dust template coefficients in flux density units. The Pearson correlations at 27 and 33\,GHz are given by $r_{27\mathrm{GHz}}$ and $r_{33\mathrm{GHz}}$, respectively.}
    \label{tab:template-coefficients-RegionB}
\renewcommand{\arraystretch}{1.5} 
\begin{threeparttable}
\begin{tabular}{cccccc|ccc}
\hline
Frequency & Wavelength & $S_{30\mathrm{GHz}}$ & Peak Frequency & $\beta$ & $T_d$ & $\alpha^{27\mathrm{GHz}}_{33\mathrm{GHz}}$ & $r_{27\mathrm{GHz}}$ & $r_{33\mathrm{GHz}}$ \\ 
      \.[GHz]&  & [$\mu$K/(MJy/sr)]   & [GHz]          &          & [K]   & & & \\ 
\hline
98.0 & 3.1mm & $29280 \pm 550$ & $33.3\pm1.8$ & $1.52\pm0.02$ & $19.1 \pm 0.3$ & $0.61 \pm 0.25$ & $0.92^{0.93}_{0.91}$ & $0.94^{0.95}_{0.93}$\\
353.0 & 0.8mm & $386 \pm 9$ & $34.7\pm2.3$ & $1.58\pm0.02$ & $18.6 \pm 0.2$ & $0.7 \pm 0.26$ & $0.90^{0.91}_{0.89}$ & $0.94^{0.95}_{0.93}$\\
545.0 & 0.6mm & $112 \pm 3$ & $35.1\pm2.5$ & $1.58\pm0.02$ & $18.6 \pm 0.2$ & $0.71 \pm 0.35$ & $0.90^{0.91}_{0.88}$ & $0.94^{0.95}_{0.93}$\\
857.0 & 0.3mm & $36.5 \pm 0.8$ & $34.4\pm2.2$ & $1.58\pm0.02$ & $18.6 \pm 0.1$ & $0.62 \pm 0.27$ & $0.90^{0.92}_{0.89}$ & $0.95^{0.96}_{0.94}$\\
3000.0 & 100$\mu$m & $34.1 \pm 0.6$ & $32.1\pm1.4$ & $1.57\pm0.03$ & $18.9 \pm 0.2$ & $0.45 \pm 0.20$ & $0.96^{0.97}_{0.95}$ & $0.96^{0.97}_{0.95}$\\
5000.0 & 60$\mu$m & $107.2 \pm 1.5$ & $31.0\pm1.1$ & $1.55\pm0.03$ & $19.1 \pm 0.2$ & $0.30 \pm 0.20$ & $0.97^{0.98}_{0.96}$ & $0.94^{0.95}_{0.93}$\\
12000.0 & 25$\mu$m & $351 \pm 4$ & $30.7\pm0.7$ & $1.55\pm0.03$ & $19.1 \pm 0.2$ & $0.28 \pm 0.12$ & $0.98^{0.99}_{0.97}$ & $0.95^{0.96}_{0.94}$\\
13600.0 & 22$\mu$m & $356 \pm 4$ & $31.1\pm0.8$ & $1.56\pm0.03$ & $19.0 \pm 0.2$ & $0.37 \pm 0.16$ & $0.98^{0.99}_{0.97}$ & $0.96^{0.97}_{0.95}$\\
25000.0 & 12$\mu$m & $444 \pm 6$ & $31.0\pm1.0$ & $1.56\pm0.04$ & $19.0 \pm 0.3$ & $0.35 \pm 0.17$ & $0.97^{0.98}_{0.96}$ & $0.95^{0.96}_{0.94}$\\
64900.0 & 4.6$\mu$m & $20030 \pm 290$ & $30.9\pm1.0$ & $1.55\pm0.03$ & $19.1 \pm 0.3$ & $0.31 \pm 0.18$ & $0.97^{0.98}_{0.96}$ & $0.94^{0.95}_{0.93}$\\
88200.0 & 3.4$\mu$m & $10770 \pm 130$ & $30.4\pm0.8$ & $1.54\pm0.03$ & $19.1 \pm 0.3$ & $0.21 \pm 0.15$ & $0.98^{0.99}_{0.97}$ & $0.94^{0.95}_{0.93}$\\\hline
\end{tabular}
\begin{tablenotes}
\item[a] The spinning dust emissivity per unit 353\,GHz dust opacity is $16.89 \pm 0.41$.
\end{tablenotes}
\end{threeparttable}
\end{table*}

\autoref{fig:template-regions} shows the COMAP 27\,GHz data after subtracting a scaled version of the GALFACTS data to remove the free-free component. The three marked regions in the figure show the pixels that were used to perform the template fitting analysis. Region \textit{Full} is a $0.\!\!^\circ5 \times 0.\!\!^\circ7$ box centred on $(\ell,b) = (192.\!\!^{\circ}3,  -11.\!\!^{\circ}6)$ contains the whole of the B30 cloud, while Region A and B are  $0.\!\!^\circ5 \times 0.\!\!^\circ35$ boxes centred on  $(\ell,b) = (192.\!\!^{\circ}3,  -11.\!\!^{\circ}425)$ and  $(\ell,b) = (192.\!\!^{\circ}3,  -11.\!\!^{\circ}775)$, respectively. The Region A and Region B pixel selections were optimised such that they could be described by a single dust tracer with a single value for the spinning dust emissivity.

\autoref{tab:template-coefficients}, \autoref{tab:template-coefficients-RegionA}, and \autoref{tab:template-coefficients-RegionB} give the best-fitting spinning dust and thermal dust model parameters for each dust tracer for the \textit{Full} region, Region A, and Region B as indicated in \autoref{fig:template-regions}. We also give the COMAP in-band spectral index, and the Pearson correlation at 27 and 33\,GHz. The best-fitting model to the IRAS\,$100$\,$\mu$m template coefficients spectrum is shown in \autoref{fig:template_spectrum}.

\subsubsection{COMAP and Dust Tracer Correlations}\label{sec:correlations}

\begin{figure}
    \centering
    \includegraphics[width=\columnwidth]{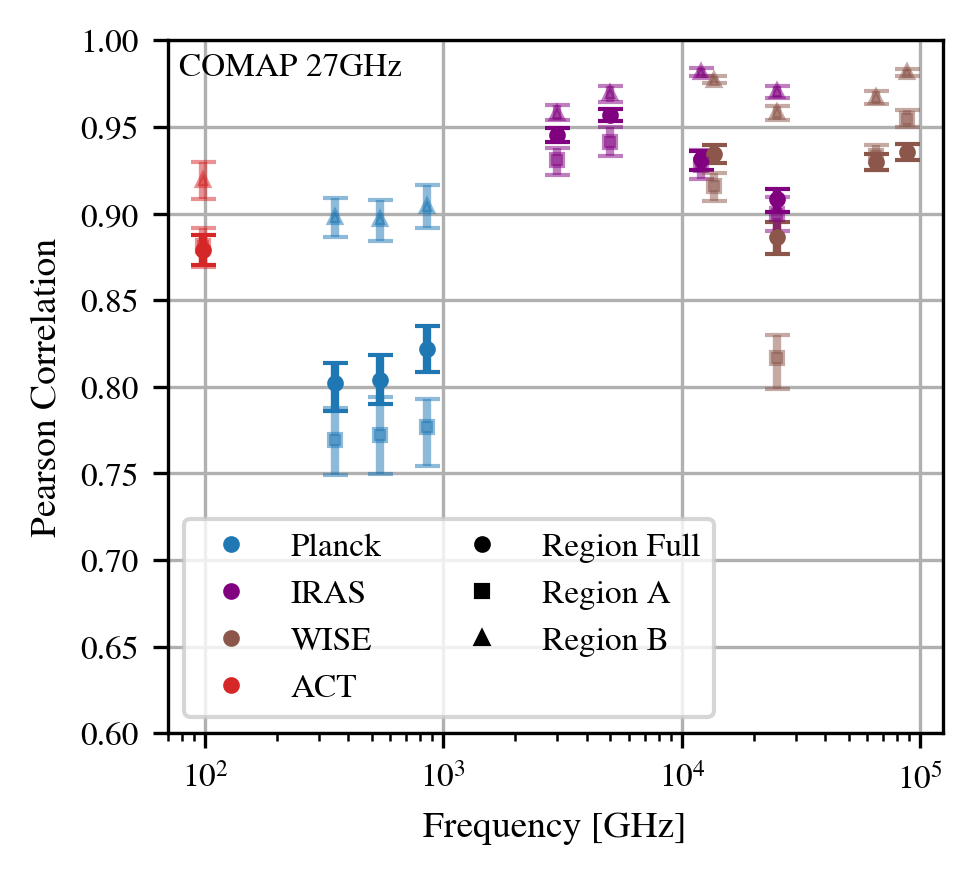}
    \includegraphics[width=\columnwidth]{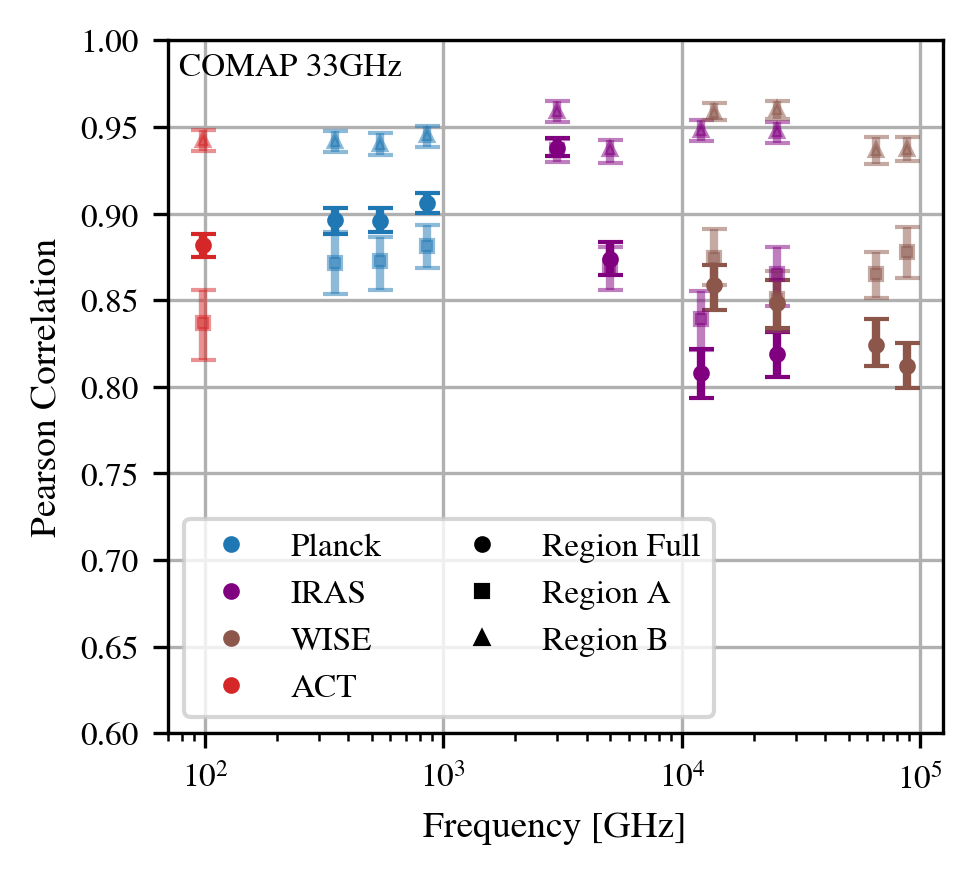}

    \caption{Pearson correlation coefficient spectrum between each dust tracer and the free-free subtracted COMAP data for Region A, B, and \textit{Full}. The \textit{top} panel shows the Pearson coefficients for the COMAP 27\,GHz data, and the \textit{bottom} panel shows the Pearson coefficients for the COMAP 33\,GHz data. }
    \label{fig:correlation-coefficients}
\end{figure}

At present the exact carrier for spinning dust is not known, and is one of the major questions facing spinning dust emission studies \citep{Dickinson2018}. For example, the \textit{Planck} sub-mm bands trace the large dust grain population in thermal equilibrium, while the FIR and MIR emission trace smaller grain populations that undergo stochastic heating events, and in the MIR there are several strong PAH features. Correlation analyses have been used successfully in the past to study spinning dust emission, but no strong conclusion has been drawn about the carrier of spinning dust emission \citep{Hensley2016,Bell2019,CepedaArroita2021}. 

In \autoref{fig:correlation-coefficients} we show the Pearson correlation coefficients between the free-free corrected COMAP 27 and 33\,GHz data and the \textit{Planck}, IRAS, and WISE maps. Both figures show the correlations measured in Region \textit{Full}, Region A, and Region B (as shown in \autoref{fig:template-regions}). If we first look at the \textit{Full} region at 27\,GHz (the \textit{circle} markers) we can see that in the sub-mm \textit{Planck} bands the correlation is lowest, it rises to a peak in the FIR around 100\,$\mu$m and then drops again in the MIR, with a slight increase in the WISE 3.4\,$\mu$m band. At 33\,GHz we see again the correlation increasing from the sub-mm to the FIR, but there is a larger drop in the MIR correlation to less than 90\,per\,cent from the 90\,per\,cent correlation seen at 27\,GHz. This difference is not simply due to the decreased sensitivity of the 33\,GHz band relative to the 27\,GHz data, as this would result in a uniform decrease in correlation with all of the dust tracers, which is not observed. The change in correlation across the band is also not due to contributions of vibrational thermal dust emission in the higher frequency COMAP bands as this is still negligible at this frequency.  Rather, this suggests that between 27 and 33\,GHz we are seeing a change in the nature of the spinning dust carrier. 

\autoref{fig:correlation-coefficients} also shows the correlation coefficients for the individual sub-regions A and B. We see that Region A is similar to the overall region correlations. The dust correlations in Region B are considerably different to those in Region A and the region overall. We can see that at 27\,GHz there is the same rise in correlation at sub-mm wavelengths, but this time the peak in the correlation is in the IRAS and WISE 25/22\,$\mu$m MIR bands, with a small fall in correlation at 12 and 4.6\,$\mu$m before rising again at 3.4\,$\mu$m. While at 33\,GHz, Region B shows no significant preference for any particular dust tracer.

How can the differences between Regions A and B be explained? One possibility is that the dust grain size distribution between Region A and B are significantly different. We know from the spinning dust model \citep{Ali-Hamoud2009} that the smallest dust grains have the highest spinning dust emissivity, and also have spectral energy distributions that peak at higher frequencies. So, if Region A has a depletion of smaller grains, we would expect to see a lower spinning dust peak frequency, and a lower spinning dust emissivity---this is exactly what we have measured.  

Using the best-fitting values to the template fitting spectrum for Region A (\autoref{tab:template-coefficients-RegionA}) and Region B (\autoref{tab:template-coefficients-RegionB}), we do find evidence that the grain size distribution and thus the spinning dust carriers change between Region A and B. First, looking at the spinning dust peak frequencies using the 100\,$\mu$m IRAS data as a dust tracer we find Region A has a peak at $27.32 \pm 0.97$\,GHz, while in Region B it peaks at $32.3 \pm 1.8$\,GHz. Second, if we use the spinning dust brightness per unit 353\,GHz optical depth as a proxy for emissivity, we find in Region A the spinning dust emissivity is $8.05\pm 0.35$\,K, while in Region B it is more than twice as high at $16.89 \pm 0.41$\,K. Both the difference in peak frequency and spinning dust emissivity match the spinning dust model predictions for a depletion of the smallest grain population. The depletion of small grains in Region A could be due to grain coagulation occurring within the known cold clump \citep{planck2015_XXVIII} and denser central cold core L1582 \citep{Kauffmann2008} within Region A. Further, previous dedicated observations of cold cores by CARMA at 30\,GHz \citep{Tibbs2011} have shown that most of these high density environments are not strong emitters of spinning dust emission.


\begin{figure}
    \centering
    \includegraphics[width=\columnwidth]{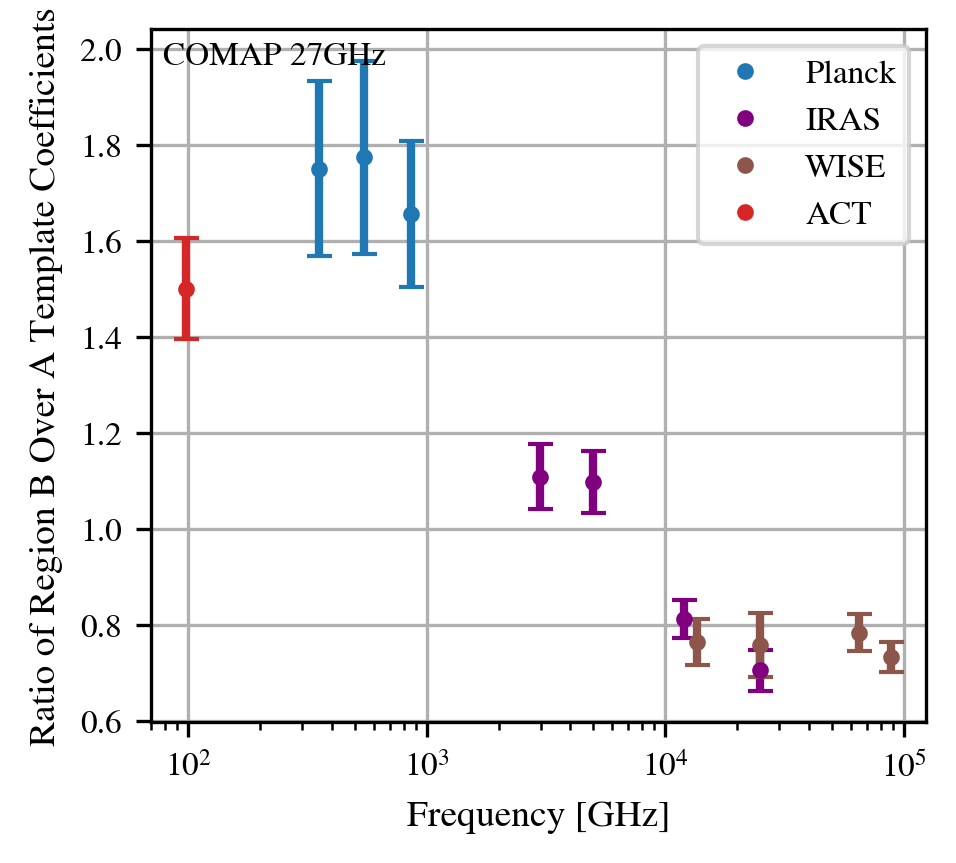}
    \caption{Ratio of best-fitting 27\,GHz template coefficients in Region B over Region A.}
    \label{fig:emissivity-ratio}
\end{figure}

In \autoref{fig:emissivity-ratio} we show how the ratio of the fitted template coefficients at 27\,GHz in Region B over A varies with dust tracer. We find that the Region B template coefficients are twice those in Region A in the sub-mm, and this ratio slowly decreases down to the MIR where the template coefficients in Region A are 20\,per\,cent brighter than those in Region B. 

The high ratio in the sub-mm emissivities between Region B and A is due to the change in the dust population grain size distribution, as discussed earlier. The smallest grains that are the most emissive spinning dust carriers are being depleted within Region A due to the high density environment of the cold clump. Such an effect has been observed several times previously, for example from the survey of cores with the CARMA telescope \citep{Tibbs2016}. Also, several studies have shown that denser environments generally have lower spinning dust emissivities relative to sub-mm bands \citep{Vidal2011,PIP_XV}, which is consistent with the idea that the small grains responsible for spinning dust emission are being depleted in Region A. 

The explanation for why the 30\,GHz emission per unit MIR brightness in Region A is apparently higher than in Region B can also be explained by the higher density environment surrounding the cold clump. In the higher density region of Region A, there will be more shielding from the incident radiation field from $\lambda$-Ori, decreasing the total ISRF. This has little effect on the emission seen in the sub-mm as these frequencies generally trace the larger grain population that are in thermal equilibrium and as such the flux density is only weakly dependent on the ISRF, e.g. $I_\nu \propto U^\frac{1}{4 + \beta}$. Meanwhile, the MIR traces grain populations that are not in thermal equilibrium, and as such are much more sensitive to the ISRF such that $I_\nu \propto U$ \citep{Draine_book}.

If the environments in Region A and B were similar, we would expect that the ratio of the flux densities in the sub-mm over the MIR should be constant. However, using the 353\,GHz and 22\,$\mu$m fluxes from \autoref{tab:flux_data}, we find that the flux density ratios for Region A and B are: $1.5 \pm 0.5$ and $0.6 \pm 0.2$ respectively. This implies that, if the MIR grains were subject to the same total radiation field in Region A as they are in B, we would expect the MIR emission to be three times higher in Region A than is observed. Therefore the apparent increase in 27\,GHz emission per unit MIR brightness in Region A is due to decrease in MIR emissivity per grain due to the decreased ISRF within Region A. 

To summarise, we find that Region A and Region B have different spinning dust template coefficients (i.e., spinning dust brightness per unit dust tracer), which is due to differences in the dust populations in both regions. The most likely explanation for why Region A has lower spinning dust template coefficients for all dust tracers is because Region A contains fewer small grains than Region B, which, from the spinning dust models \citep{Ali-Hamoud2009}, is expected to reduce the spinning dust emissivity and peak frequency. However, we cannot entirely rule out that these differences are due to other factors such as differences in environment or dust grain dipole moments. If there are fewer small dust grains in Region A, one possible explanation for this is that the small grains are coagulating within the cold clump region resulting in the small grains being depleted. To confirm this we will need spectroscopic MIR observations to find direct evidence of small grain depletion. Potentially this could be done with the James Webb Space Telescopes (JWST) MIR instrument similar to recent JWST observations of the Orion bar \citep{Berne2022}. 

\subsubsection{Comparisons with Other Work}

There have been several other studies of the correlation between spinning dust emission and dust tracers around the $\lambda$-Orionis PDR at $1^\circ$ resolution \citep{Bell2019,CepedaArroita2021,Chuss2022}. These studies looked at the correlation between maps of spinning dust of the entire PDR derived from per pixel parametric fitting of multiple $1^\circ$ resolution surveys. Both lower resolution studies of the $\lambda$-Orionis PDR found that spinning dust best correlated with 857\,GHz data, with a Spearman correlation of $r_\mathrm{857GHz} \approx 0.9$ and a sharp decline in correlation in the FIR, reaching $r_{60\mu\mathrm{m}} \approx 0.4$. 

We do not observe a sharp decline in the correlation around 60\,$\mu$m for B30, but in fact find that that the spinning dust correlation peaks in the FIR. Both \citet{Bell2019} and \citet{CepedaArroita2021} attribute the drop in correlation at 60\,$\mu$m to stochastically heated small grains, that are not in thermal equilibrium, not being good carriers of spinning dust emission. However, this is not true for B30. A potential reason is that the spinning dust emissivity is varying significantly around the $\lambda$-Orionis PDR due to variations in environment and dust grain size distribution. It has been noted for some time that the spinning dust emissivity relative to FIR data can exhibit large variations due to environment \citep{Tibbs2012b}. We see this effect when we measure Region \textit{Full}, as shown in \autoref{fig:correlation-coefficients}, there is a sharp drop in the FIR/MIR at 33\,GHz when measuring Region \textit{Full}. We also tried measuring larger regions, and found that the correlation for FIR/MIR decreases as you increase the area fitted by the template fitting analysis. While noise could be responsible for the observed decrease in correlation, another possibility is variations in the spinning dust emissivity. Thus the results of the low resolution analysis may be driven by large variations in spinning dust emissivity, which is more sensitive at FIR/MIR bands as these bands are more sensitive to environmental factors such as the ISRF and dust temperature. 

\subsubsection{Template Fitting Residuals}

\begin{figure*}
    \includegraphics[width=\textwidth]{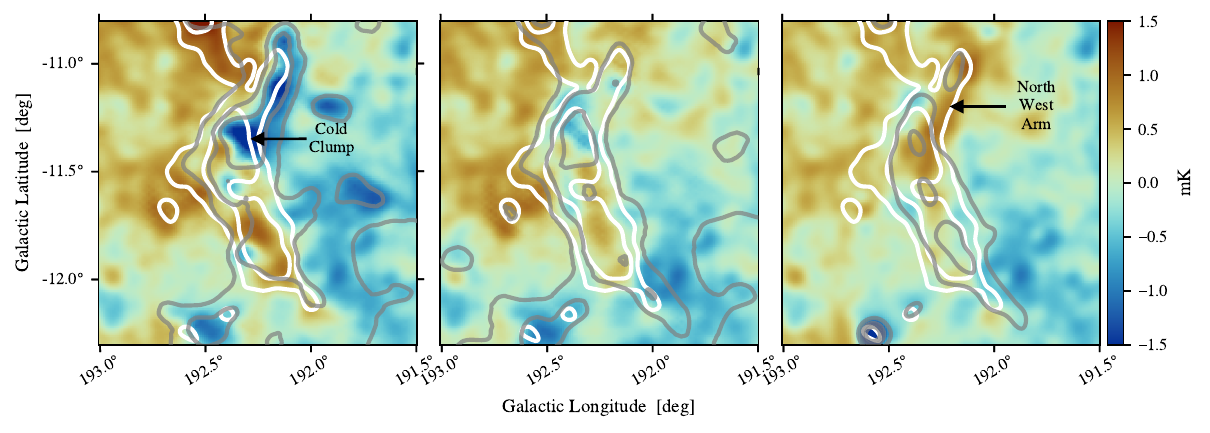}
    \caption{27\,GHz residuals after subtracting the free-free background and best-fitting template fitted dust model using the Region B template coefficients. Dust maps subtracted, from \textit{left} to \textit{right} are: \textit{Planck} 353\,GHz, IRAS 100\,$\mu$m, and WISE 22\,$\mu$m. The \textit{white} contours show the COMAP 27\,GHz data, while the \textit{gray} contours show the morphology of the dust tracers. Each panel has dimensions of $1.\!\!^\circ5 \times 1.\!\!^\circ5$.}\label{fig:template-residuals}
\end{figure*}

In \autoref{fig:template-residuals} we show the residual 27\,GHz data after subtracting the free-free background using GALFACTS and the best-fitting template fitted dust model using \textit{Planck} 353\,GHz, IRAS 100\,$\mu$m, and WISE 22\,$\mu$m. We have used the template coefficients fitted to Region B to highlight the impact of the cold clump and the changes in an extension we have labelled on the WISE 22\,$\mu$m map as the North West Arm. 

We can see from the residual using the \textit{Planck} 353\,GHz data that the sub-mm data are not a good tracer of the 27\,GHz emission. There is both a positive residual in Region B, indicating a poor fit, and that in the region around both the Cold Clump and the North West Arm in Region A, there is a clear negative or over-subtraction. Additionally, there is a gradient running from \textit{left} to \textit{right} across the image, with a positive residual that is tracing the edge of the free-free emission seen at 1.4\,GHz. We are seeing this effect as the sub-mm data is insensitive to the total radiation field that is driven by $\lambda$-Orionis, which implies that the spinning dust emission per grain is higher in regions towards the irradiated edge of the PDR, and decreases in either more shielded regions (i.e., within the cold clump), or in regions farther away from $\lambda$-Orionis.

In the the \textit{right}-most panel of \autoref{fig:template-residuals} we show the residual after subtracting the WISE 22\,$\mu$m data. In this panel we can see that the residual in Region B is much closer to zero. However, the residuals around the cold clump and the North West arm are positive, i.e., the spinning dust emission per unit MIR brightness is higher in Region A, which is what we also see in \autoref{fig:emissivity-ratio} showing the ratio of the template coefficients in Region B over A. This further supports the earlier discussion that the MIR grains are emitting less in Region A (and the North West arm feature) due to the increased dust column density and the resulting decrease in the ISRF in this region.

The \textit{centre} panel shows the FIR IRAS 100\,$\mu$m residual. In this residual map we can see that the residuals are much closer to zero across the whole field, with a small excess in Region B, and small over subtraction around the cold clump. We also see the gradient from \textit{left} to \textit{right} that we see in all three residuals. FIR emission is sensitive to both the total radiation field and the dust column along a line-of-sight, and has only a small contribution from stochastically heated grains. Interestingly, spinning dust models do not predict any significant change in spinning dust emissivity with dust temperature \citep{Ali-Hamoud2009}, so the high correlation with FIR tracers, which are very sensitive to dust temperature, is surprising. However, this suggests that the depletion of the spinning dust carriers in regions with lower temperature is directly proportional to the decrease in FIR emission due to the dust temperature. Unfortunately, without spectroscopic observations of the dust population in B30, it hard to make a definitive statement about relationship between dust grain population and temperature. 

Finally, the North West arm is seen in both the \textit{Planck} 353\,GHz as a negative, and the WISE 22\,$\mu$m data as a positive residual. This feature can be seen in all of the maps shown in \autoref{fig:grid}, and its location relative to COMAP shifts from being to the \textit{right} of the COMAP data in the \textit{Planck} sub-mm bands, to aligning with COMAP in the FIR, and then to the \textit{left} of COMAP in the MIR. This is a classic feature of a PDR, as different grain size populations emit in different infrared bands, with smaller grains emitting in the MIR in regions with higher total radiation fields, and larger grains emitting in the FIR/sub-mm as the grains become more shielded. The fact the FIR residual in the North West arm is consistent with zero in \autoref{fig:template-residuals}, is a strong indicator that the spinning dust emission is predominantly associated with the dust population that is also emitting in the FIR. The fact that this region is also well modelled by the template coefficient fitted to Region B, suggests that the same dust population is emitting in both the North West arm and Region B.

\subsubsection{Spinning Dust Emissivity and Spectrum}

A way to determine the origin of spinning dust within a region is to look at the spinning dust emissivity defined as the spinning dust emission brightness at a given frequency per unit dust tracer. Common dust tracers that are used include the IRAS 100\,$\mu$m map, the optical depth at 353\,GHz, the total integrated dust radiance, and MIR data containing PAH features \citep{Dickinson2018}. Each of these are tracers of different dust populations, and are susceptible to different environmental conditions. 

We compared the ratios of spinning dust emission to the dust optical depth at 353\,GHz and to emission at 100\,$\mu$m for Region A and Region B to ratios from the literature. We find that Region B is typical of most spinning dust emitting regions, with a spinning dust brightness per unit 353\,GHz optical depth of $16.89 \pm 0.41$\,K, and a ratio to $I_{100\mu\mathrm{m}}$ of $34.11 \pm 0.58$\,$\mu$K/(MJy/sr). For reference the Perseus and $\rho$-Ophiuchus clouds have ratios of $16.4 \pm 3.1$\,K, and $24.3 \pm 4.0$\,$\mu$K/(MJy/sr) and $14.2 \pm 4.0$\,K, and $9.6 \pm 2.1$\,$\mu$K/(MJy/sr), respectively \citep{Planck2011_XX}. Both of these regions are star forming regions and are associated with PDRs. Comparing the ratios of the Region B template coefficients to those of Perseus and $\rho$-Ophiuchus, we find that per unit $\tau_{353}$ (i.e., the total dust column) these regions are very similar, however for FIR $I_{100\mu\mathrm{m}}$ emission the ratio of Region B to Perseus is 1.2, while for $\rho$-Ophiuchus it is 3.6. It is not entirely unexpected that there is a larger scatter in the relative brightness to FIR tracers, as far FIR emission is more dependent on the local radiation field \citep{Tibbs2012b}. However, this does imply that, since the emission per grain is the same in all of these PDRs, that spinning dust emission can originate in a range of environmental conditions and is not sensitive to the range of environmental differences between these regions. 

We find that the peak frequency of spinning dust is around 30\,GHz for the whole cloud, with a peak frequency slightly below 30\,GHz in Region A, and slightly above 30\,GHz in Region B. We have limited data to constrain the peak of the spinning dust spectrum in B30 (data at 15\,GHz and 50 to 60\,GHz would greatly help), however we can use the COMAP in-band spectral indices, given in \autoref{tab:template-coefficients-RegionA} and \autoref{tab:template-coefficients-RegionB}, to see that for Region A the in-band spectral index is consistent with zero implying we are near to the peak. For Region B, the spectrum is rising with a significance between 2 and 3$\sigma$, depending on the dust template, which is what is driving the slightly higher peak frequency---however, we cannot rule out a higher peak frequency in Region B entirely with COMAP alone. The peak frequency of both Region A and B are consistent with typical spinning dust peak frequencies associated with PDRs (e.g., Perseus and $\rho$-Ophiuchus \citep{Planck2011_XX}), but also the typical peak frequencies from surveys of spinning dust emitting regions within the Galaxy that are likely associated with PDR environments \citep{PIP_XV,Tramonte2023}.

The most well-studied source of spinning dust emission however comes from high-latitude spinning dust clouds, where strong correlations between between microwave and FIR data has been well-known since the earliest AME studies \citep{Banday2003,Davies2006}. The best-fitting template coefficients of these high-latitude clouds are $14.0 \pm 0.4$\,$\mu$K/(MJy/sr) relative to $I_{100\mu\mathrm{m}}$, and $5.8\pm 0.2$\,K relative to $\tau_{353}$ \citep{Harper2023}, which are a factor of 2.4 and 2.9 lower than the template coefficients measured in Region B, and 1.4 and 2.4 times lower than in Region A. Clearly this suggests that the dust grain population and environment in high-latitude cirrus clouds is very different to that in B30 and other PDR regions. The lower emissivity and peak frequency in the high-latitude clouds suggests that there is a depletion of smaller grain sizes in these regions, similar to what we are finding in Region A. Spinning dust emission in high-latitude clouds have also been shown to not strongly correlate with PAH fraction \citep{Hensley2016,Hensley2022}. Considering that both environments can produce spinning dust emission, and that we see no strong preference that the COMAP data correlate with MIR bands that have PAH feature contamination, this suggests that PAHs are not required for spinning dust emission and may not be the primary spinning dust carrier.

There have also been a number of measurements of AME regions that have defined the spinning dust emission per unit IRAS 100\,$\mu$m brightness at resolutions comparable to COMAP. For example, six H\textsc{ii} regions measured by the Cosmic Background Interferometer (CBI) at 31\,GHz found average emissivities of $3.3 \pm 1.7$\,$\mu$K/(MJy/sr), except for RCW49 that had an emissivity of $13.6\pm4.2$ \citep{Dickinson2007}. Nine other H\textsc{ii} regions surveyed by \citep{Todorovic2010} found an average emissivity of $3.9\pm 0.8$\,$\mu$K/(MJy/sr). The CBI was also used to measure the emissivity of two Lynds dark clouds LDN\,1622 \citep{Casassus2006} and LDN\,1621 \citep{Dickinson2010}, and found emissivities of $21.3\pm0.6$ and $18.1 \pm 4.4$\,$\mu$K/(MJy/sr), respectively. Finally, the Perseus molecular cloud, when measured by high resolution observations tends to yield emissivities that are significantly lower than those found by \textit{Planck} at low resolutions. In summary, all of these values are much lower than the emissivity of B30 in either Region A or B by 50\,per\,cent or more. However, the emissivity that we measure for B30 is consistent with that found by \textit{Planck} at $1^\circ$ resolution. The main difference between COMAP and previous high-resolution measurements of spinning dust regions is that all previous observations used interferometric observations, while COMAP is a single-dish total power telescope. This means COMAP does not filter out extended emission, and so the difference in spinning dust emissivity indicates that a significant proportion of the total spinning dust emission originates in the more diffuse, extended phases of the ISM.

\section{Conclusions}\label{sec:conclusions}

B30 is a sub-region of the PDR bounding the edge of the $\lambda$-Orionis H\textsc{ii} region. The B30 region has been studied several times at $1^\circ$ resolution \citep{PIP_XV,Bell2019,CepedaArroita2021} and was identified as a promising spinning dust emitting candidate. In this paper we have presented $4.\!^\prime5$ resolution observations from COMAP, which, combined with ancillary data from GALFACTS, ACT, \textit{Planck}, IRAS, and WISE, have revealed B30 to be one of the cleanest examples of spinning dust detected to date.

We have studied B30, and sub-regions within B30, using both aperture photometry and template fitting and find that $86.0 \pm 0.08$\,per\,cent of the emission at 30\,GHz is due to spinning dust, which we detect above the free-free background with $7\sigma$ significance. Using template fitting we have identified that the spinning dust emission is best traced with the FIR and MIR data at wavelengths less than 100\,$\mu$m, with a weak preference for the 60\,$\mu$m band. However, we find there is no preference for MIR bands that contain PAH features, raising further doubts about an association between spinning dust emission and PAHs.

We have also found evidence of variations of spinning dust emissivity within B30. Region B has a higher spinning dust emissivity per 353\,GHz optical depth than Region A, when using the dust optical depth as a tracer, while Region A is slightly brighter per unit MIR brightness. We suggest that the difference in emissivity per dust grain in Region A is due to differences in the dust grain population in both regions. We suggest that the most likely explanation is that Region A has a smaller population of the smallest dust grains than Region B, which is predicted by spinning dust models to result in lower spinning dust emissivity and peak frequency just as we have measured in B30 \citep{Ali-Hamoud2009}. A possible reason for the depletion of small grains in Region A could be because of coagulation of small grains in to larger grains within the high density cold clump within Region A. However, we cannot rule out that the differences in the spinning dust emissivity between Region A and B are not due to other effects, such as: differences grain dipole moments, or environmental factors. Certainly we know that environmental factors must be making an impact as we find the MIR and FIR emissivities are also different in Region A and B, with Region A having less MIR emission per unit dust column that is likely due to a lower total ISRF due to increased shielding from the higher density environment. Ultimately, to fully understand the physics of what is happening within B30 will require spectroscopic observations to quantify the dust grain population. This could be achieved using JWST MIRI observations similar to those recently undertaken of the Orion bar \citep{Berne2022}. 


To conclude, these new observations of B30 within the $\lambda$-Orionis ring by the COMAP Pathfinder are revealing new possibilities in understanding the physics of spinning dust emission. Further studies of spinning dust regions with COMAP are planned in the near future by extending the Galactic plane survey presented in \citet{Rennie2022}, as well as further targets including the Perseus molecular cloud, the Andromeda galaxy, and the California nebula. Finally, these observations highlight the critical importance of single-dish, rather than interferometric, observations to measure spinning dust emission. This is because we find most of the spinning dust emission is in the diffuse ISM, and not compact sources, which can be filtered out by interferometric observations. As for B30, we intend to revisit this analysis in the future using planned 15\,GHz COMAP observations \citep{Cleary2022,Breysse2022}, which will allow for more detailed modelling of the spinning dust spectrum and lead to a more definitive answer to the question of what is the carrier of spinning dust in B30.

\section*{Acknowledgements}

SEH and CD acknowledge funding from an STFC Consolidated Grant (ST/P000649/1) and a UKSA grant (ST/Y005945/1) funding LiteBIRD foreground activities.

This material is based upon work supported by the National Science Foundation under Grant Nos.\ 1517108, 1517288, 1517598, 1518282, 1910999 and 2206834, as well as by the Keck Institute for Space Studies under ``The First Billion Years: A Technical Development Program for Spectral Line Observations''.

This work was carried out in part at Jet Propulsion Laboratory, California Institute of Technology, under a contract with the National Aeronautics and Space Administration.

We acknowledge support from the Research Council of Norway through grants 251328 and 274990, and from the European Research Council (ERC) under the Horizon 2020 Research and Innovation Program (Grant agreement No. 772253 bits2cosmology and 819478 Cosmoglobe).

\section*{Data Availability} 

The COMAP 27, 29, 31, and 33\,GHz maps of Barnard~30 used in this analysis can be made available upon request to the authors.

\bibliographystyle{mnras}
\bibliography{cbassbiblio}

\begin{thebibliography}{}
\makeatletter
\relax
\def\mn@urlcharsother{\let\do\@makeother \do\$\do\&\do\#\do\^\do\_\do\%\do\~}
\def\mn@doi{\begingroup\mn@urlcharsother \@ifnextchar [ {\mn@doi@}
  {\mn@doi@[]}}
\def\mn@doi@[#1]#2{\def\@tempa{#1}\ifx\@tempa\@empty \href
  {http://dx.doi.org/#2} {doi:#2}\else \href {http://dx.doi.org/#2} {#1}\fi
  \endgroup}
\def\mn@eprint#1#2{\mn@eprint@#1:#2::\@nil}
\def\mn@eprint@arXiv#1{\href {http://arxiv.org/abs/#1} {{\tt arXiv:#1}}}
\def\mn@eprint@dblp#1{\href {http://dblp.uni-trier.de/rec/bibtex/#1.xml}
  {dblp:#1}}
\def\mn@eprint@#1:#2:#3:#4\@nil{\def\@tempa {#1}\def\@tempb {#2}\def\@tempc
  {#3}\ifx \@tempc \@empty \let \@tempc \@tempb \let \@tempb \@tempa \fi \ifx
  \@tempb \@empty \def\@tempb {arXiv}\fi \@ifundefined
  {mn@eprint@\@tempb}{\@tempb:\@tempc}{\expandafter \expandafter \csname
  mn@eprint@\@tempb\endcsname \expandafter{\@tempc}}}

\bibitem[\protect\citeauthoryear{{Abdo} et~al.,}{{Abdo}
  et~al.}{2010}]{Abdo2010}
{Abdo} A.~A.,  et~al., 2010, \mn@doi [\apj] {10.1088/0004-637X/716/1/30}, \href
  {https://ui.adsabs.harvard.edu/abs/2010ApJ...716...30A} {716, 30}

\bibitem[\protect\citeauthoryear{{Ackermann} et~al.,}{{Ackermann}
  et~al.}{2011}]{Ackermann2011}
{Ackermann} M.,  et~al., 2011, \mn@doi [\apj] {10.1088/0004-637X/741/1/30},
  \href {https://ui.adsabs.harvard.edu/abs/2011ApJ...741...30A} {741, 30}

\bibitem[\protect\citeauthoryear{{Ali-Ha{\"\i}moud}, {Hirata}  \&
  {Dickinson}}{{Ali-Ha{\"\i}moud} et~al.}{2009}]{Ali-Hamoud2009}
{Ali-Ha{\"\i}moud} Y.,  {Hirata} C.~M.,   {Dickinson} C.,  2009, \mn@doi
  [\mnras] {10.1111/j.1365-2966.2009.14599.x}, \href
  {http://adsabs.harvard.edu/abs/2009MNRAS.395.1055A} {395, 1055}

\bibitem[\protect\citeauthoryear{{Banday}, {Dickinson}, {Davies}, {Davis}  \&
  {G{\'o}rski}}{{Banday} et~al.}{2003}]{Banday2003}
{Banday} A.~J.,  {Dickinson} C.,  {Davies} R.~D.,  {Davis} R.~J.,
  {G{\'o}rski} K.~M.,  2003, \mn@doi [\mnras]
  {10.1046/j.1365-8711.2003.07008.x}, \href
  {http://adsabs.harvard.edu/abs/2003MNRAS.345..897B} {345, 897}

\bibitem[\protect\citeauthoryear{{Barrado} et~al.,}{{Barrado}
  et~al.}{2018}]{Barrado2018}
{Barrado} D.,  et~al., 2018, \mn@doi [\aap] {10.1051/0004-6361/201527938},
  \href {https://ui.adsabs.harvard.edu/abs/2018A&A...612A..79B} {612, A79}

\bibitem[\protect\citeauthoryear{{Becker}, {White}  \& {Edwards}}{{Becker}
  et~al.}{1991}]{Becker1991}
{Becker} R.~H.,  {White} R.~L.,   {Edwards} A.~L.,  1991, \mn@doi [\apjs]
  {10.1086/191529}, \href
  {https://ui.adsabs.harvard.edu/abs/1991ApJS...75....1B} {75, 1}

\bibitem[\protect\citeauthoryear{{Bell}, {Onaka}, {Galliano}, {Wu}, {Doi},
  {Kaneda}, {Ishihara}  \& {Giard}}{{Bell} et~al.}{2019}]{Bell2019}
{Bell} A.~C.,  {Onaka} T.,  {Galliano} F.,  {Wu} R.,  {Doi} Y.,  {Kaneda} H.,
  {Ishihara} D.,   {Giard} M.,  2019, \mn@doi [\pasj] {10.1093/pasj/psz110},
  \href {https://ui.adsabs.harvard.edu/abs/2019PASJ..tmp..110B} {p.~110}

\bibitem[\protect\citeauthoryear{{Bennett}, {Lawrence}, {Burke}, {Hewitt}  \&
  {Mahoney}}{{Bennett} et~al.}{1986}]{Bennett1986}
{Bennett} C.~L.,  {Lawrence} C.~R.,  {Burke} B.~F.,  {Hewitt} J.~N.,
  {Mahoney} J.,  1986, \mn@doi [\apjs] {10.1086/191108}, \href
  {https://ui.adsabs.harvard.edu/abs/1986ApJS...61....1B} {61, 1}

\bibitem[\protect\citeauthoryear{{Bern{\'e}} et~al.,}{{Bern{\'e}}
  et~al.}{2022}]{Berne2022}
{Bern{\'e}} O.,  et~al., 2022, \mn@doi [\pasp] {10.1088/1538-3873/ac604c},
  \href {https://ui.adsabs.harvard.edu/abs/2022PASP..134e4301B} {134, 054301}

\bibitem[\protect\citeauthoryear{{Breysse} et~al.,}{{Breysse}
  et~al.}{2022}]{Breysse2022}
{Breysse} P.~C.,  et~al., 2022, \mn@doi [\apj] {10.3847/1538-4357/ac63c9},
  \href {https://ui.adsabs.harvard.edu/abs/2022ApJ...933..188B} {933, 188}

\bibitem[\protect\citeauthoryear{{Cara} \& {Lister}}{{Cara} \&
  {Lister}}{2008}]{Cara2008}
{Cara} M.,  {Lister} M.~L.,  2008, \mn@doi [\apj] {10.1086/525554}, \href
  {https://ui.adsabs.harvard.edu/abs/2008ApJ...674..111C} {674, 111}

\bibitem[\protect\citeauthoryear{{Casassus}, {Cabrera}, {F{\"o}rster},
  {Pearson}, {Readhead}  \& {Dickinson}}{{Casassus}
  et~al.}{2006}]{Casassus2006}
{Casassus} S.,  {Cabrera} G.~F.,  {F{\"o}rster} F.,  {Pearson} T.~J.,
  {Readhead} A.~C.~S.,   {Dickinson} C.,  2006, \mn@doi [\apj]
  {10.1086/499517}, \href {http://adsabs.harvard.edu/abs/2006ApJ...639..951C}
  {639, 951}

\bibitem[\protect\citeauthoryear{{Casassus} et~al.,}{{Casassus}
  et~al.}{2008}]{Casassus2008}
{Casassus} S.,  et~al., 2008, \mn@doi [\mnras]
  {10.1111/j.1365-2966.2008.13954.x}, \href
  {http://adsabs.harvard.edu/abs/2008MNRAS.391.1075C} {391, 1075}

\bibitem[\protect\citeauthoryear{{Cepeda-Arroita} et~al.,}{{Cepeda-Arroita}
  et~al.}{2021}]{CepedaArroita2021}
{Cepeda-Arroita} R.,  et~al., 2021, \mn@doi [\mnras] {10.1093/mnras/stab583},
  \href {https://ui.adsabs.harvard.edu/abs/2021MNRAS.503.2927C} {503, 2927}

\bibitem[\protect\citeauthoryear{{Chuss}, {Hensley}, {Kogut}, {Guerra}, {Nofi}
  \& {Siah}}{{Chuss} et~al.}{2022}]{Chuss2022}
{Chuss} D.~T.,  {Hensley} B.~S.,  {Kogut} A.~J.,  {Guerra} J.~A.,  {Nofi}
  H.~C.,   {Siah} J.,  2022, \mn@doi [\apj] {10.3847/1538-4357/ac9b24}, \href
  {https://ui.adsabs.harvard.edu/abs/2022ApJ...940...59C} {940, 59}

\bibitem[\protect\citeauthoryear{{Cleary} et~al.,}{{Cleary}
  et~al.}{2022}]{Cleary2022}
{Cleary} K.~A.,  et~al., 2022, \mn@doi [\apj] {10.3847/1538-4357/ac63cc}, \href
  {https://ui.adsabs.harvard.edu/abs/2022ApJ...933..182C} {933, 182}

\bibitem[\protect\citeauthoryear{{Condon}, {Cotton}, {Greisen}, {Yin},
  {Perley}, {Taylor}  \& {Broderick}}{{Condon} et~al.}{1998}]{Condon1998}
{Condon} J.~J.,  {Cotton} W.~D.,  {Greisen} E.~W.,  {Yin} Q.~F.,  {Perley}
  R.~A.,  {Taylor} G.~B.,   {Broderick} J.~J.,  1998, \mn@doi [\aj]
  {10.1086/300337}, \href {http://adsabs.harvard.edu/abs/1998AJ....115.1693C}
  {115, 1693}

\bibitem[\protect\citeauthoryear{{Cooper}, {Lister}  \& {Kochanczyk}}{{Cooper}
  et~al.}{2007}]{Cooper2007}
{Cooper} N.~J.,  {Lister} M.~L.,   {Kochanczyk} M.~D.,  2007, \mn@doi [\apjs]
  {10.1086/518654}, \href
  {https://ui.adsabs.harvard.edu/abs/2007ApJS..171..376C} {171, 376}

\bibitem[\protect\citeauthoryear{{Cutri} et~al.,}{{Cutri}
  et~al.}{2012}]{Cutri2012}
{Cutri} R.~M.,  et~al., 2012, {Explanatory Supplement to the WISE All-Sky Data
  Release Products}, Explanatory Supplement to the WISE All-Sky Data Release
  Products

\bibitem[\protect\citeauthoryear{{Davies}, {Dickinson}, {Banday}, {Jaffe},
  {G{\'o}rski}  \& {Davis}}{{Davies} et~al.}{2006}]{Davies2006}
{Davies} R.~D.,  {Dickinson} C.,  {Banday} A.~J.,  {Jaffe} T.~R.,  {G{\'o}rski}
  K.~M.,   {Davis} R.~J.,  2006, \mn@doi [\mnras]
  {10.1111/j.1365-2966.2006.10572.x}, \href
  {http://adsabs.harvard.edu/abs/2006MNRAS.370.1125D} {370, 1125}

\bibitem[\protect\citeauthoryear{{Delabrouille}}{{Delabrouille}}{1998}]{Delabrouille1998}
{Delabrouille} J.,  1998, \mn@doi [\aaps] {10.1051/aas:1998119}, \href
  {https://ui.adsabs.harvard.edu/abs/1998A&AS..127..555D} {127, 555}

\bibitem[\protect\citeauthoryear{{Dennison}, {Simonetti}  \&
  {Topasna}}{{Dennison} et~al.}{1998}]{Dennison1998}
{Dennison} B.,  {Simonetti} J.~H.,   {Topasna} G.~A.,  1998, Publications
  Astronomical Society of Australia, \href
  {http://adsabs.harvard.edu/abs/1998PASA...15..147D} {15, 147}

\bibitem[\protect\citeauthoryear{{Dickinson}, {Davies}  \& {Davis}}{{Dickinson}
  et~al.}{2003}]{Dickinson2003}
{Dickinson} C.,  {Davies} R.~D.,   {Davis} R.~J.,  2003, \mn@doi [\mnras]
  {10.1046/j.1365-8711.2003.06439.x}, \href
  {http://adsabs.harvard.edu/abs/2003MNRAS.341..369D} {341, 369}

\bibitem[\protect\citeauthoryear{{Dickinson}, {Davies}, {Bronfman}, {Casassus},
  {Davis}, {Pearson}, {Readhead}  \& {Wilkinson}}{{Dickinson}
  et~al.}{2007}]{Dickinson2007}
{Dickinson} C.,  {Davies} R.~D.,  {Bronfman} L.,  {Casassus} S.,  {Davis}
  R.~J.,  {Pearson} T.~J.,  {Readhead} A.~C.~S.,   {Wilkinson} P.~N.,  2007,
  \mn@doi [\mnras] {10.1111/j.1365-2966.2007.11967.x}, \href
  {http://adsabs.harvard.edu/abs/2007MNRAS.379..297D} {379, 297}

\bibitem[\protect\citeauthoryear{{Dickinson} et~al.,}{{Dickinson}
  et~al.}{2010}]{Dickinson2010}
{Dickinson} C.,  et~al., 2010, \mn@doi [\mnras]
  {10.1111/j.1365-2966.2010.17079.x}, \href
  {http://adsabs.harvard.edu/abs/2010MNRAS.407.2223D} {407, 2223}

\bibitem[\protect\citeauthoryear{{Dickinson} et~al.,}{{Dickinson}
  et~al.}{2018}]{Dickinson2018}
{Dickinson} C.,  et~al., 2018, \mn@doi [\nar] {10.1016/j.newar.2018.02.001},
  \href {http://adsabs.harvard.edu/abs/2018NewAR..80....1D} {80, 1}

\bibitem[\protect\citeauthoryear{{Dodson} et~al.,}{{Dodson}
  et~al.}{2008}]{Dodson2008}
{Dodson} R.,  et~al., 2008, \mn@doi [\apjs] {10.1086/525025}, \href
  {https://ui.adsabs.harvard.edu/abs/2008ApJS..175..314D} {175, 314}

\bibitem[\protect\citeauthoryear{{Dolan} \& {Mathieu}}{{Dolan} \&
  {Mathieu}}{2001}]{Dolan2001}
{Dolan} C.~J.,  {Mathieu} R.~D.,  2001, \mn@doi [\aj] {10.1086/319946}, \href
  {https://ui.adsabs.harvard.edu/abs/2001AJ....121.2124D} {121, 2124}

\bibitem[\protect\citeauthoryear{{Draine}}{{Draine}}{2011}]{Draine_book}
{Draine} B.~T.,  2011, {Physics of the Interstellar and Intergalactic Medium}.
Princeton University Press

\bibitem[\protect\citeauthoryear{{Draine} \& {Lazarian}}{{Draine} \&
  {Lazarian}}{1998}]{Draine1998a}
{Draine} B.~T.,  {Lazarian} A.,  1998, \mn@doi [\apjl] {10.1086/311167}, \href
  {http://adsabs.harvard.edu/abs/1998ApJ...494L..19D} {494, L19}

\bibitem[\protect\citeauthoryear{{Draine} \& {Lazarian}}{{Draine} \&
  {Lazarian}}{1999}]{Draine1999}
{Draine} B.~T.,  {Lazarian} A.,  1999, \mn@doi [\apj] {10.1086/306809}, \href
  {http://adsabs.harvard.edu/abs/1999ApJ...512..740D} {512, 740}

\bibitem[\protect\citeauthoryear{{Foreman-Mackey}, {Hogg}, {Lang}  \&
  {Goodman}}{{Foreman-Mackey} et~al.}{2013}]{Foreman-Mackey2013}
{Foreman-Mackey} D.,  {Hogg} D.~W.,  {Lang} D.,   {Goodman} J.,  2013, \mn@doi
  [\pasp] {10.1086/670067}, \href
  {http://adsabs.harvard.edu/abs/2013PASP..125..306F} {125, 306}

\bibitem[\protect\citeauthoryear{{G{\'o}rski}, {Hivon}, {Banday}, {Wandelt},
  {Hansen}, {Reinecke}  \& {Bartelmann}}{{G{\'o}rski}
  et~al.}{2005}]{Gorski2005}
{G{\'o}rski} K.~M.,  {Hivon} E.,  {Banday} A.~J.,  {Wandelt} B.~D.,  {Hansen}
  F.~K.,  {Reinecke} M.,   {Bartelmann} M.,  2005, \mn@doi [\apj]
  {10.1086/427976}, \href {http://adsabs.harvard.edu/abs/2005ApJ...622..759G}
  {622, 759}

\bibitem[\protect\citeauthoryear{{Greaves}, {Scaife}, {Frayer}, {Green},
  {Mason}  \& {Smith}}{{Greaves} et~al.}{2018}]{Greaves2018}
{Greaves} J.~S.,  {Scaife} A.~M.~M.,  {Frayer} D.~T.,  {Green} D.~A.,  {Mason}
  B.~S.,   {Smith} A.~M.~S.,  2018, \mn@doi [Nature Astronomy]
  {10.1038/s41550-018-0495-z}, \href
  {https://ui.adsabs.harvard.edu/abs/2018NatAs...2..662G} {2, 662}

\bibitem[\protect\citeauthoryear{{Gregory} \& {Condon}}{{Gregory} \&
  {Condon}}{1991}]{Gregory1991}
{Gregory} P.~C.,  {Condon} J.~J.,  1991, \mn@doi [\apjs] {10.1086/191559},
  \href {https://ui.adsabs.harvard.edu/abs/1991ApJS...75.1011G} {75, 1011}

\bibitem[\protect\citeauthoryear{{Gregory}, {Scott}, {Douglas}  \&
  {Condon}}{{Gregory} et~al.}{1996}]{Gregory1996}
{Gregory} P.~C.,  {Scott} W.~K.,  {Douglas} K.,   {Condon} J.~J.,  1996,
  \mn@doi [\apjs] {10.1086/192282}, \href
  {http://adsabs.harvard.edu/abs/1996ApJS..103..427G} {103, 427}

\bibitem[\protect\citeauthoryear{{Harper}}{{Harper}}{2024}]{Harper2024}
{Harper} S. E. e.~a.,  2024, \mnras, in prep.

\bibitem[\protect\citeauthoryear{{Harper}, {Dickinson}  \& {Cleary}}{{Harper}
  et~al.}{2015}]{Harper2015}
{Harper} S.~E.,  {Dickinson} C.,   {Cleary} K.,  2015, \mn@doi [\mnras]
  {10.1093/mnras/stv1863}, \href
  {https://ui.adsabs.harvard.edu/abs/2015MNRAS.453.3375H} {453, 3375}

\bibitem[\protect\citeauthoryear{{Harper} et~al.,}{{Harper}
  et~al.}{2022}]{Harper2022}
{Harper} S.~E.,  et~al., 2022, \mn@doi [\mnras] {10.1093/mnras/stac1210}, \href
  {https://ui.adsabs.harvard.edu/abs/2022MNRAS.513.5900H} {513, 5900}

\bibitem[\protect\citeauthoryear{{Harper} et~al.,}{{Harper}
  et~al.}{2023}]{Harper2023}
{Harper} S.~E.,  et~al., 2023, \mn@doi [\mnras] {10.1093/mnras/stad1539}, \href
  {https://ui.adsabs.harvard.edu/abs/2023MNRAS.523.3471H} {523, 3471}

\bibitem[\protect\citeauthoryear{{Healey}, {Romani}, {Taylor}, {Sadler},
  {Ricci}, {Murphy}, {Ulvestad}  \& {Winn}}{{Healey} et~al.}{2007}]{Healey2007}
{Healey} S.~E.,  {Romani} R.~W.,  {Taylor} G.~B.,  {Sadler} E.~M.,  {Ricci} R.,
   {Murphy} T.,  {Ulvestad} J.~S.,   {Winn} J.~N.,  2007, \mn@doi [\apjs]
  {10.1086/513742}, \href
  {https://ui.adsabs.harvard.edu/abs/2007ApJS..171...61H} {171, 61}

\bibitem[\protect\citeauthoryear{{Healey}, {Fuhrmann}, {Taylor}, {Romani}  \&
  {Readhead}}{{Healey} et~al.}{2009}]{Healey2009}
{Healey} S.~E.,  {Fuhrmann} L.,  {Taylor} G.~B.,  {Romani} R.~W.,   {Readhead}
  A.~C.~S.,  2009, \mn@doi [\aj] {10.1088/0004-6256/138/4/1032}, \href
  {http://adsabs.harvard.edu/abs/2009AJ....138.1032H} {138, 1032}

\bibitem[\protect\citeauthoryear{{Hensley}, {Murphy}  \& {Staguhn}}{{Hensley}
  et~al.}{2015}]{Hensley2015}
{Hensley} B.,  {Murphy} E.,   {Staguhn} J.,  2015, \mn@doi [\mnras]
  {10.1093/mnras/stv287}, \href
  {https://ui.adsabs.harvard.edu/abs/2015MNRAS.449..809H} {449, 809}

\bibitem[\protect\citeauthoryear{{Hensley}, {Draine}  \& {Meisner}}{{Hensley}
  et~al.}{2016}]{Hensley2016}
{Hensley} B.~S.,  {Draine} B.~T.,   {Meisner} A.~M.,  2016, \mn@doi [\apj]
  {10.3847/0004-637X/827/1/45}, \href
  {http://adsabs.harvard.edu/abs/2016ApJ...827...45H} {827, 45}

\bibitem[\protect\citeauthoryear{{Hensley}, {Murray}  \& {Dodici}}{{Hensley}
  et~al.}{2022}]{Hensley2022}
{Hensley} B.~S.,  {Murray} C.~E.,   {Dodici} M.,  2022, \mn@doi [\apj]
  {10.3847/1538-4357/ac5cbd}, \href
  {https://ui.adsabs.harvard.edu/abs/2022ApJ...929...23H} {929, 23}

\bibitem[\protect\citeauthoryear{{Hilbe}, {de Souza}  \& {Ishida}}{{Hilbe}
  et~al.}{2017}]{Hilbe2017}
{Hilbe} J.~M.,  {de Souza} R.~S.,   {Ishida} E. E.~O.,  2017, {Bayesian Models
  for Astrophysical Data Using R, JAGS, Python, and Stan}.
Cambridge University Press, \mn@doi{10.1017/CBO9781316459515}

\bibitem[\protect\citeauthoryear{{Hoang} \& {Lazarian}}{{Hoang} \&
  {Lazarian}}{2016}]{Hoang2016}
{Hoang} T.,  {Lazarian} A.,  2016, \mn@doi [\apj] {10.3847/0004-637X/821/2/91},
  \href {http://adsabs.harvard.edu/abs/2016ApJ...821...91H} {821, 91}

\bibitem[\protect\citeauthoryear{{Horiuchi} et~al.,}{{Horiuchi}
  et~al.}{2004}]{Horiuchi2004}
{Horiuchi} S.,  et~al., 2004, \mn@doi [\apj] {10.1086/424811}, \href
  {https://ui.adsabs.harvard.edu/abs/2004ApJ...616..110H} {616, 110}

\bibitem[\protect\citeauthoryear{{Hovatta}, {Nieppola}, {Tornikoski},
  {Valtaoja}, {Aller}  \& {Aller}}{{Hovatta} et~al.}{2008}]{Hovatta2008}
{Hovatta} T.,  {Nieppola} E.,  {Tornikoski} M.,  {Valtaoja} E.,  {Aller} M.~F.,
    {Aller} H.~D.,  2008, \mn@doi [\aap] {10.1051/0004-6361:200809806}, \href
  {https://ui.adsabs.harvard.edu/abs/2008A&A...485...51H} {485, 51}

\bibitem[\protect\citeauthoryear{{Hu{\'e}lamo} et~al.,}{{Hu{\'e}lamo}
  et~al.}{2017}]{Huelamo2017}
{Hu{\'e}lamo} N.,  et~al., 2017, \mn@doi [\aap] {10.1051/0004-6361/201628510},
  \href {https://ui.adsabs.harvard.edu/abs/2017A&A...597A..17H} {597, A17}

\bibitem[\protect\citeauthoryear{{Kauffmann}, {Bertoldi}, {Bourke}, {Evans}  \&
  {Lee}}{{Kauffmann} et~al.}{2008}]{Kauffmann2008}
{Kauffmann} J.,  {Bertoldi} F.,  {Bourke} T.~L.,  {Evans} N.~J. I.,   {Lee}
  C.~W.,  2008, \mn@doi [\aap] {10.1051/0004-6361:200809481}, \href
  {https://ui.adsabs.harvard.edu/abs/2008A&A...487..993K} {487, 993}

\bibitem[\protect\citeauthoryear{{Kellermann} et~al.,}{{Kellermann}
  et~al.}{2004}]{Kellermann2004}
{Kellermann} K.~I.,  et~al., 2004, \mn@doi [\apj] {10.1086/421289}, \href
  {https://ui.adsabs.harvard.edu/abs/2004ApJ...609..539K} {609, 539}

\bibitem[\protect\citeauthoryear{{Koay} et~al.,}{{Koay}
  et~al.}{2011}]{Koay2011}
{Koay} J.~Y.,  et~al., 2011, \mn@doi [\aj] {10.1088/0004-6256/142/4/108}, \href
  {https://ui.adsabs.harvard.edu/abs/2011AJ....142..108K} {142, 108}

\bibitem[\protect\citeauthoryear{{Koenig}, {Hillenbrand}, {Padgett}  \&
  {DeFelippis}}{{Koenig} et~al.}{2015}]{Koenig2015}
{Koenig} X.,  {Hillenbrand} L.~A.,  {Padgett} D.~L.,   {DeFelippis} D.,  2015,
  \mn@doi [\aj] {10.1088/0004-6256/150/4/100}, \href
  {https://ui.adsabs.harvard.edu/abs/2015AJ....150..100K} {150, 100}

\bibitem[\protect\citeauthoryear{{Kogut}, {Banday}, {Bennett}, {Gorski},
  {Hinshaw}, {Smoot}  \& {Wright}}{{Kogut} et~al.}{1996}]{Kogut1996}
{Kogut} A.,  {Banday} A.~J.,  {Bennett} C.~L.,  {Gorski} K.~M.,  {Hinshaw} G.,
  {Smoot} G.~F.,   {Wright} E.~I.,  1996, \mn@doi [\apjl] {10.1086/310072},
  \href {http://ukads.nottingham.ac.uk/abs/1996ApJ...464L...5K} {464, L5}

\bibitem[\protect\citeauthoryear{{Kovalev} et~al.,}{{Kovalev}
  et~al.}{2005}]{Kovalev2005}
{Kovalev} Y.~Y.,  et~al., 2005, \mn@doi [\aj] {10.1086/497430}, \href
  {https://ui.adsabs.harvard.edu/abs/2005AJ....130.2473K} {130, 2473}

\bibitem[\protect\citeauthoryear{{Kuehr}, {Witzel}, {Pauliny-Toth}  \&
  {Nauber}}{{Kuehr} et~al.}{1981}]{Kuehr1981}
{Kuehr} H.,  {Witzel} A.,  {Pauliny-Toth} I.~I.~K.,   {Nauber} U.,  1981,
  \aaps, \href {https://ui.adsabs.harvard.edu/abs/1981A&AS...45..367K} {45,
  367}

\bibitem[\protect\citeauthoryear{{Lamb} et~al.,}{{Lamb}
  et~al.}{2022}]{Lamb2022}
{Lamb} J.~W.,  et~al., 2022, \mn@doi [\apj] {10.3847/1538-4357/ac63c6}, \href
  {https://ui.adsabs.harvard.edu/abs/2022ApJ...933..183L} {933, 183}

\bibitem[\protect\citeauthoryear{{Lang}, {Masheder}, {Dame}  \&
  {Thaddeus}}{{Lang} et~al.}{2000}]{Lang2000}
{Lang} W.~J.,  {Masheder} M.~R.~W.,  {Dame} T.~M.,   {Thaddeus} P.,  2000,
  \aap, \href {http://adsabs.harvard.edu/abs/2000A%26A...357.1001L} {357, 1001}

\bibitem[\protect\citeauthoryear{{Laurent-Muehleisen}, {Kollgaard}, {Ryan},
  {Feigelson}, {Brinkmann}  \& {Siebert}}{{Laurent-Muehleisen}
  et~al.}{1997}]{Laurent-Muehleisen1997}
{Laurent-Muehleisen} S.~A.,  {Kollgaard} R.~I.,  {Ryan} P.~J.,  {Feigelson}
  E.~D.,  {Brinkmann} W.,   {Siebert} J.,  1997, \mn@doi [\aaps]
  {10.1051/aas:1997331}, \href
  {https://ui.adsabs.harvard.edu/abs/1997A&AS..122..235L} {122, 235}

\bibitem[\protect\citeauthoryear{{Lee}, {Sohn}, {Jung}, {Byun}  \& {Lee}}{{Lee}
  et~al.}{2017}]{Jeong2017}
{Lee} J.~A.,  {Sohn} B.~W.,  {Jung} T.,  {Byun} D.-Y.,   {Lee} J.~W.,  2017,
  \mn@doi [\apjs] {10.3847/1538-4365/228/2/22}, \href
  {https://ui.adsabs.harvard.edu/abs/2017ApJS..228...22L} {228, 22}

\bibitem[\protect\citeauthoryear{{Leighton}}{{Leighton}}{1977}]{Leighton1977}
{Leighton} R.~B.,  1977, {Millimeter-wave antenna design}, Final Report
  California Inst. of Tech., Pasadena.

\bibitem[\protect\citeauthoryear{{Leitch}, {Readhead}, {Pearson}  \&
  {Myers}}{{Leitch} et~al.}{1997}]{Leitch1997}
{Leitch} E.~M.,  {Readhead} A.~C.~S.,  {Pearson} T.~J.,   {Myers} S.~T.,  1997,
  \mn@doi [\apjl] {10.1086/310823}, \href
  {http://adsabs.harvard.edu/abs/1997ApJ...486L..23L} {486, L23}

\bibitem[\protect\citeauthoryear{{Linford}, {Taylor}, {Romani}, {Helmboldt},
  {Readhead}, {Reeves}  \& {Richards}}{{Linford} et~al.}{2012}]{Linford2012}
{Linford} J.~D.,  {Taylor} G.~B.,  {Romani} R.~W.,  {Helmboldt} J.~F.,
  {Readhead} A.~C.~S.,  {Reeves} R.,   {Richards} J.~L.,  2012, \mn@doi [\apj]
  {10.1088/0004-637X/744/2/177}, \href
  {https://ui.adsabs.harvard.edu/abs/2012ApJ...744..177L} {744, 177}

\bibitem[\protect\citeauthoryear{{Lister} et~al.,}{{Lister}
  et~al.}{2011}]{Lister2011}
{Lister} M.~L.,  et~al., 2011, \mn@doi [\apj] {10.1088/0004-637X/742/1/27},
  \href {https://ui.adsabs.harvard.edu/abs/2011ApJ...742...27L} {742, 27}

\bibitem[\protect\citeauthoryear{{Lister}, {Aller}, {Aller}, {Hovatta},
  {Max-Moerbeck}, {Readhead}, {Richards}  \& {Ros}}{{Lister}
  et~al.}{2015}]{Lister2015}
{Lister} M.~L.,  {Aller} M.~F.,  {Aller} H.~D.,  {Hovatta} T.,  {Max-Moerbeck}
  W.,  {Readhead} A.~C.~S.,  {Richards} J.~L.,   {Ros} E.,  2015, \mn@doi
  [\apjl] {10.1088/2041-8205/810/1/L9}, \href
  {https://ui.adsabs.harvard.edu/abs/2015ApJ...810L...9L} {810, L9}

\bibitem[\protect\citeauthoryear{{Maddalena} \& {Morris}}{{Maddalena} \&
  {Morris}}{1987}]{Maddalena1987}
{Maddalena} R.~J.,  {Morris} M.,  1987, \mn@doi [\apj] {10.1086/165818}, \href
  {http://adsabs.harvard.edu/abs/1987ApJ...323..179M} {323, 179}

\bibitem[\protect\citeauthoryear{{Massardi}, {L{\'o}pez-Caniego},
  {Gonz{\'a}lez-Nuevo}, {Herranz}, {de Zotti}  \& {Sanz}}{{Massardi}
  et~al.}{2009}]{Massardi2009}
{Massardi} M.,  {L{\'o}pez-Caniego} M.,  {Gonz{\'a}lez-Nuevo} J.,  {Herranz}
  D.,  {de Zotti} G.,   {Sanz} J.~L.,  2009, \mn@doi [\mnras]
  {10.1111/j.1365-2966.2008.14084.x}, \href
  {https://ui.adsabs.harvard.edu/abs/2009MNRAS.392..733M} {392, 733}

\bibitem[\protect\citeauthoryear{{Miville-Desch{\^e}nes} \&
  {Lagache}}{{Miville-Desch{\^e}nes} \&
  {Lagache}}{2005}]{Miville-Deschenes2005}
{Miville-Desch{\^e}nes} M.,  {Lagache} G.,  2005, \mn@doi [\apjs]
  {10.1086/427938}, \href {http://adsabs.harvard.edu/abs/2005ApJS..157..302M}
  {157, 302}

\bibitem[\protect\citeauthoryear{{Murdin} \& {Penston}}{{Murdin} \&
  {Penston}}{1977}]{Murdin1977}
{Murdin} P.,  {Penston} M.~V.,  1977, \mn@doi [\mnras]
  {10.1093/mnras/181.4.657}, \href
  {http://adsabs.harvard.edu/abs/1977MNRAS.181..657M} {181, 657}

\bibitem[\protect\citeauthoryear{{Murphy} et~al.,}{{Murphy}
  et~al.}{2010}]{Murphy2010}
{Murphy} E.~J.,  et~al., 2010, \mn@doi [\apjl] {10.1088/2041-8205/709/2/L108},
  \href {http://adsabs.harvard.edu/abs/2010ApJ...709L.108M} {709, L108}

\bibitem[\protect\citeauthoryear{{Naess} et~al.,}{{Naess}
  et~al.}{2020}]{Naess2020}
{Naess} S.,  et~al., 2020, \mn@doi [\jcap] {10.1088/1475-7516/2020/12/046},
  \href {https://ui.adsabs.harvard.edu/abs/2020JCAP...12..046N} {2020, 046}

\bibitem[\protect\citeauthoryear{{Neugebauer} et~al.,}{{Neugebauer}
  et~al.}{1984}]{Neugebauer1984}
{Neugebauer} G.,  et~al., 1984, \mn@doi [\apjl] {10.1086/184209}, \href
  {https://ui.adsabs.harvard.edu/abs/1984ApJ...278L...1N} {278, L1}

\bibitem[\protect\citeauthoryear{{Nieppola}, {Tornikoski}, {Valtaoja},
  {Le{\'o}n-Tavares}, {Hovatta}, {L{\"a}hteenm{\"a}ki}  \& {Tammi}}{{Nieppola}
  et~al.}{2011}]{Nieppola2011}
{Nieppola} E.,  {Tornikoski} M.,  {Valtaoja} E.,  {Le{\'o}n-Tavares} J.,
  {Hovatta} T.,  {L{\"a}hteenm{\"a}ki} A.,   {Tammi} J.,  2011, \mn@doi [\aap]
  {10.1051/0004-6361/201116818}, \href
  {https://ui.adsabs.harvard.edu/abs/2011A&A...535A..69N} {535, A69}

\bibitem[\protect\citeauthoryear{{Paladini}, {Burigana}, {Davies}, {Maino},
  {Bersanelli}, {Cappellini}, {Platania}  \& {Smoot}}{{Paladini}
  et~al.}{2003}]{Paladini2003}
{Paladini} R.,  {Burigana} C.,  {Davies} R.~D.,  {Maino} D.,  {Bersanelli} M.,
  {Cappellini} B.,  {Platania} P.,   {Smoot} G.,  2003, \mn@doi [\aap]
  {10.1051/0004-6361:20021466}, \href
  {https://ui.adsabs.harvard.edu/abs/2003A&A...397..213P} {397, 213}

\bibitem[\protect\citeauthoryear{{Parsons} et~al.,}{{Parsons}
  et~al.}{2008}]{Parsons2008}
{Parsons} A.,  et~al., 2008, \mn@doi [\pasp] {10.1086/593053}, \href
  {https://ui.adsabs.harvard.edu/abs/2008PASP..120.1207P} {120, 1207}

\bibitem[\protect\citeauthoryear{{Peel}, {Dickinson}, {Davies}, {Banday},
  {Jaffe}  \& {Jonas}}{{Peel} et~al.}{2012}]{Peel2012}
{Peel} M.~W.,  {Dickinson} C.,  {Davies} R.~D.,  {Banday} A.~J.,  {Jaffe}
  T.~R.,   {Jonas} J.~L.,  2012, \mn@doi [\mnras]
  {10.1111/j.1365-2966.2012.21358.x}, \href
  {http://ukads.nottingham.ac.uk/abs/2012MNRAS.424.2676P} {424, 2676}

\bibitem[\protect\citeauthoryear{{Petrov}, {Hirota}, {Honma}, {Shibata}, {Jike}
   \& {Kobayashi}}{{Petrov} et~al.}{2007}]{Petrov2007}
{Petrov} L.,  {Hirota} T.,  {Honma} M.,  {Shibata} K.~M.,  {Jike} T.,
  {Kobayashi} H.,  2007, \mn@doi [\aj] {10.1086/513146}, \href
  {https://ui.adsabs.harvard.edu/abs/2007AJ....133.2487P} {133, 2487}

\bibitem[\protect\citeauthoryear{{Planck Collaboration} et~al.,}{{Planck
  Collaboration} et~al.}{2011a}]{Planck2011_XV}
{Planck Collaboration} et~al., 2011a, \mn@doi [\aap]
  {10.1051/0004-6361/201116466}, \href
  {https://ui.adsabs.harvard.edu/abs/2011A&A...536A..15P} {536, A15}

\bibitem[\protect\citeauthoryear{{Planck Collaboration} et~al.,}{{Planck
  Collaboration} et~al.}{2011b}]{Planck2011_XX}
{Planck Collaboration} et~al., 2011b, \mn@doi [\aap]
  {10.1051/0004-6361/201116470}, \href
  {https://ui.adsabs.harvard.edu/\#abs/2011A&A...536A..20P} {536, A20}

\bibitem[\protect\citeauthoryear{{Planck Collaboration} et~al.,}{{Planck
  Collaboration} et~al.}{2014a}]{PIP_XV}
{Planck Collaboration} et~al., 2014a, \mn@doi [\aap]
  {10.1051/0004-6361/201322612}, \href
  {http://adsabs.harvard.edu/abs/2014A%26A...565A.103P} {565, A103}

\bibitem[\protect\citeauthoryear{{Planck Collaboration} et~al.,}{{Planck
  Collaboration} et~al.}{2014b}]{Planck2013_XI}
{Planck Collaboration} et~al., 2014b, \mn@doi [\aap]
  {10.1051/0004-6361/201323195}, \href
  {http://adsabs.harvard.edu/abs/2014A%26A...571A..11P} {571, A11}

\bibitem[\protect\citeauthoryear{{Planck Collaboration} et~al.,}{{Planck
  Collaboration} et~al.}{2016a}]{Planck2016_XXIX}
{Planck Collaboration} et~al., 2016a, \mn@doi [\aap]
  {10.1051/0004-6361/201424945}, \href
  {https://ui.adsabs.harvard.edu/abs/2016A&A...586A.132P} {586, A132}

\bibitem[\protect\citeauthoryear{{Planck Collaboration} et~al.,}{{Planck
  Collaboration} et~al.}{2016b}]{Planck2015_X}
{Planck Collaboration} et~al., 2016b, \mn@doi [\aap]
  {10.1051/0004-6361/201525967}, \href
  {https://ui.adsabs.harvard.edu/\#abs/2016A&A...594A..10P} {594, A10}

\bibitem[\protect\citeauthoryear{{Planck Collaboration} et~al.,}{{Planck
  Collaboration} et~al.}{2016c}]{planck2015_XXVIII}
{Planck Collaboration} et~al., 2016c, \mn@doi [\aap]
  {10.1051/0004-6361/201525819}, \href
  {https://ui.adsabs.harvard.edu/abs/2016A&A...594A..28P} {594, A28}

\bibitem[\protect\citeauthoryear{{Planck Collaboration} et~al.,}{{Planck
  Collaboration} et~al.}{2020}]{Planck2018_I}
{Planck Collaboration} et~al., 2020, \mn@doi [\aap]
  {10.1051/0004-6361/201833880}, \href
  {https://ui.adsabs.harvard.edu/abs/2020A&A...641A...1P} {641, A1}

\bibitem[\protect\citeauthoryear{{Poidevin} et~al.,}{{Poidevin}
  et~al.}{2023}]{Poidevin2023}
{Poidevin} F.,  et~al., 2023, \mn@doi [\mnras] {10.1093/mnras/stac3151}, \href
  {https://ui.adsabs.harvard.edu/abs/2023MNRAS.519.3481P} {519, 3481}

\bibitem[\protect\citeauthoryear{{Rennie} et~al.,}{{Rennie}
  et~al.}{2022}]{Rennie2022}
{Rennie} T.~J.,  et~al., 2022, \mn@doi [\apj] {10.3847/1538-4357/ac63c8}, \href
  {https://ui.adsabs.harvard.edu/abs/2022ApJ...933..187R} {933, 187}

\bibitem[\protect\citeauthoryear{{Richards}, {Hovatta}, {Max-Moerbeck},
  {Pavlidou}, {Pearson}  \& {Readhead}}{{Richards} et~al.}{2014}]{Richards2014}
{Richards} J.~L.,  {Hovatta} T.,  {Max-Moerbeck} W.,  {Pavlidou} V.,  {Pearson}
  T.~J.,   {Readhead} A.~C.~S.,  2014, \mn@doi [\mnras]
  {10.1093/mnras/stt2412}, \href
  {https://ui.adsabs.harvard.edu/abs/2014MNRAS.438.3058R} {438, 3058}

\bibitem[\protect\citeauthoryear{{Rickett}, {Lazio}  \& {Ghigo}}{{Rickett}
  et~al.}{2006}]{Rickett2006}
{Rickett} B.~J.,  {Lazio} T.~J.~W.,   {Ghigo} F.~D.,  2006, \mn@doi [\apjs]
  {10.1086/504897}, \href
  {https://ui.adsabs.harvard.edu/abs/2006ApJS..165..439R} {165, 439}

\bibitem[\protect\citeauthoryear{{Sharpless}}{{Sharpless}}{1959}]{Sharpless1959}
{Sharpless} S.,  1959, \mn@doi [\apjs] {10.1086/190049}, \href
  {http://adsabs.harvard.edu/abs/1959ApJS....4..257S} {4, 257}

\bibitem[\protect\citeauthoryear{{Shimmins}, {Bolton}  \& {Wall}}{{Shimmins}
  et~al.}{1975}]{Shimmins1975}
{Shimmins} A.~J.,  {Bolton} J.~G.,   {Wall} J.~V.,  1975, Australian Journal of
  Physics Astrophysical Supplement, \href
  {https://ui.adsabs.harvard.edu/abs/1975AuJPA..34...63S} {34, 63}

\bibitem[\protect\citeauthoryear{{Silsbee}, {Ali-Ha{\"\i}moud}  \&
  {Hirata}}{{Silsbee} et~al.}{2011}]{Silsbee2011}
{Silsbee} K.,  {Ali-Ha{\"\i}moud} Y.,   {Hirata} C.~M.,  2011, \mn@doi [\mnras]
  {10.1111/j.1365-2966.2010.17882.x}, \href
  {http://adsabs.harvard.edu/abs/2011MNRAS.411.2750S} {411, 2750}

\bibitem[\protect\citeauthoryear{{Stanghellini}, {O'Dea}, {Dallacasa}, {Baum},
  {Fanti}  \& {Fanti}}{{Stanghellini} et~al.}{1998}]{Stanghellini1998}
{Stanghellini} C.,  {O'Dea} C.~P.,  {Dallacasa} D.,  {Baum} S.~A.,  {Fanti} R.,
    {Fanti} C.,  1998, \mn@doi [\aaps] {10.1051/aas:1998270}, \href
  {https://ui.adsabs.harvard.edu/abs/1998A&AS..131..303S} {131, 303}

\bibitem[\protect\citeauthoryear{{Sutton} et~al.,}{{Sutton}
  et~al.}{2010}]{Sutton2010}
{Sutton} D.,  et~al., 2010, \mn@doi [\mnras]
  {10.1111/j.1365-2966.2010.16954.x}, \href
  {http://adsabs.harvard.edu/abs/2010MNRAS.407.1387S} {407, 1387}

\bibitem[\protect\citeauthoryear{Taylor}{Taylor}{2013}]{Taylor2013}
Taylor A.~R.,  2013, \mn@doi [IOP Conference Series: Materials Science and
  Engineering] {10.1088/1757-899X/44/1/012019}, 44, 012019

\bibitem[\protect\citeauthoryear{{Tibbs} et~al.,}{{Tibbs}
  et~al.}{2011}]{Tibbs2011}
{Tibbs} C.~T.,  et~al., 2011, \mn@doi [\mnras]
  {10.1111/j.1365-2966.2011.19605.x}, \href
  {http://adsabs.harvard.edu/abs/2011MNRAS.418.1889T} {418, 1889}

\bibitem[\protect\citeauthoryear{{Tibbs} et~al.,}{{Tibbs}
  et~al.}{2012a}]{Tibbs2012}
{Tibbs} C.~T.,  et~al., 2012a, \mn@doi [\apj] {10.1088/0004-637X/754/2/94},
  \href {http://adsabs.harvard.edu/abs/2012ApJ...754...94T} {754, 94}

\bibitem[\protect\citeauthoryear{{Tibbs}, {Paladini}  \& {Dickinson}}{{Tibbs}
  et~al.}{2012b}]{Tibbs2012b}
{Tibbs} C.~T.,  {Paladini} R.,   {Dickinson} C.,  2012b, \mn@doi [Advances in
  Astronomy] {10.1155/2012/124931}, \href
  {http://adsabs.harvard.edu/abs/2012AdAst2012E..41T} {2012, 124931}

\bibitem[\protect\citeauthoryear{{Tibbs} et~al.,}{{Tibbs}
  et~al.}{2016}]{Tibbs2016}
{Tibbs} C.~T.,  et~al., 2016, \mn@doi [\mnras] {10.1093/mnras/stv2759}, \href
  {https://ui.adsabs.harvard.edu/abs/2016MNRAS.456.2290T} {456, 2290}

\bibitem[\protect\citeauthoryear{{Tielens}}{{Tielens}}{2008}]{Tielens2008}
{Tielens} A.~G.~G.~M.,  2008, \mn@doi [\araa]
  {10.1146/annurev.astro.46.060407.145211}, \href
  {http://adsabs.harvard.edu/abs/2008ARA%26A..46..289T} {46, 289}

\bibitem[\protect\citeauthoryear{{Todorovi{\'c}} et~al.,}{{Todorovi{\'c}}
  et~al.}{2010}]{Todorovic2010}
{Todorovi{\'c}} M.,  et~al., 2010, \mn@doi [\mnras]
  {10.1111/j.1365-2966.2010.16809.x}, \href
  {http://adsabs.harvard.edu/abs/2010MNRAS.406.1629T} {406, 1629}

\bibitem[\protect\citeauthoryear{{Tramonte} et~al.,}{{Tramonte}
  et~al.}{2023}]{Tramonte2023}
{Tramonte} D.,  et~al., 2023, \mn@doi [\mnras] {10.1093/mnras/stac3502}, \href
  {https://ui.adsabs.harvard.edu/abs/2023MNRAS.519.3432T} {519, 3432}

\bibitem[\protect\citeauthoryear{{Vidal} et~al.,}{{Vidal}
  et~al.}{2011}]{Vidal2011}
{Vidal} M.,  et~al., 2011, \mn@doi [\mnras] {10.1111/j.1365-2966.2011.18562.x},
  \href {http://adsabs.harvard.edu/abs/2011MNRAS.414.2424V} {414, 2424}

\bibitem[\protect\citeauthoryear{{Vidal}, {Dickinson}, {Harper}, {Casassus}  \&
  {Witt}}{{Vidal} et~al.}{2020}]{Vidal2019}
{Vidal} M.,  {Dickinson} C.,  {Harper} S.~E.,  {Casassus} S.,   {Witt} A.~N.,
  2020, \mn@doi [\mnras] {10.1093/mnras/staa1186}, \href
  {https://ui.adsabs.harvard.edu/abs/2020MNRAS.495.1122V} {495, 1122}

\bibitem[\protect\citeauthoryear{{Vollmer}, {Krichbaum}, {Angelakis}  \&
  {Kovalev}}{{Vollmer} et~al.}{2008}]{Vollmer2008}
{Vollmer} B.,  {Krichbaum} T.~P.,  {Angelakis} E.,   {Kovalev} Y.~Y.,  2008,
  \mn@doi [\aap] {10.1051/0004-6361:20078857}, \href
  {https://ui.adsabs.harvard.edu/abs/2008A&A...489...49V} {489, 49}

\bibitem[\protect\citeauthoryear{{Weiland} et~al.,}{{Weiland}
  et~al.}{2011}]{Weiland2011}
{Weiland} J.~L.,  et~al., 2011, \mn@doi [\apjs] {10.1088/0067-0049/192/2/19},
  \href {http://adsabs.harvard.edu/abs/2011ApJS..192...19W} {192, 19}

\bibitem[\protect\citeauthoryear{{White} \& {Becker}}{{White} \&
  {Becker}}{1992}]{White1992a}
{White} R.~L.,  {Becker} R.~H.,  1992, \mn@doi [\apjs] {10.1086/191656}, \href
  {https://ui.adsabs.harvard.edu/abs/1992ApJS...79..331W} {79, 331}

\bibitem[\protect\citeauthoryear{{Wiren}, {Valtaoja}, {Terasranta}  \&
  {Kotilainen}}{{Wiren} et~al.}{1992}]{Wiren1992}
{Wiren} S.,  {Valtaoja} E.,  {Terasranta} H.,   {Kotilainen} J.,  1992, \mn@doi
  [\aj] {10.1086/116294}, \href
  {https://ui.adsabs.harvard.edu/abs/1992AJ....104.1009W} {104, 1009}

\bibitem[\protect\citeauthoryear{{Wright} \& {Otrupcek}}{{Wright} \&
  {Otrupcek}}{1990}]{Wright1990}
{Wright} A.,  {Otrupcek} R.,  1990, PKS Catalog (1990, \href
  {https://ui.adsabs.harvard.edu/abs/1990PKS...C......0W} {p.~0}

\bibitem[\protect\citeauthoryear{{Wright} et~al.,}{{Wright}
  et~al.}{2010}]{Wright2010WISE}
{Wright} E.~L.,  et~al., 2010, \mn@doi [\aj] {10.1088/0004-6256/140/6/1868},
  \href {https://ui.adsabs.harvard.edu/\#abs/2010AJ....140.1868W} {140, 1868}

\bibitem[\protect\citeauthoryear{{Ysard}, {Miville-Desch{\^e}nes}, {Verstraete}
   \& {Jones}}{{Ysard} et~al.}{2022}]{Ysard2022}
{Ysard} N.,  {Miville-Desch{\^e}nes} M.~A.,  {Verstraete} L.,   {Jones} A.~P.,
  2022, \mn@doi [\aap] {10.1051/0004-6361/202142825}, \href
  {https://ui.adsabs.harvard.edu/abs/2022A&A...663A..65Y} {663, A65}

\bibitem[\protect\citeauthoryear{{Zavala} \& {Taylor}}{{Zavala} \&
  {Taylor}}{2004}]{Zavala2004}
{Zavala} R.~T.,  {Taylor} G.~B.,  2004, \mn@doi [\apj] {10.1086/422741}, \href
  {https://ui.adsabs.harvard.edu/abs/2004ApJ...612..749Z} {612, 749}

\bibitem[\protect\citeauthoryear{{de Oliveira-Costa}, {Kogut}, {Devlin},
  {Netterfield}, {Page}  \& {Wollack}}{{de Oliveira-Costa}
  et~al.}{1997}]{deOliveira-Costa1997}
{de Oliveira-Costa} A.,  {Kogut} A.,  {Devlin} M.~J.,  {Netterfield} C.~B.,
  {Page} L.~A.,   {Wollack} E.~J.,  1997, \mn@doi [\apjl] {10.1086/310684},
  \href {http://adsabs.harvard.edu/abs/1997ApJ...482L..17D} {482, L17}

\makeatother
\end{thebibliography}

\appendix

\section{QSO~J0530+13}\label{sec:qso}

\begin{figure*}
    \centering
    \includegraphics[width=\textwidth]{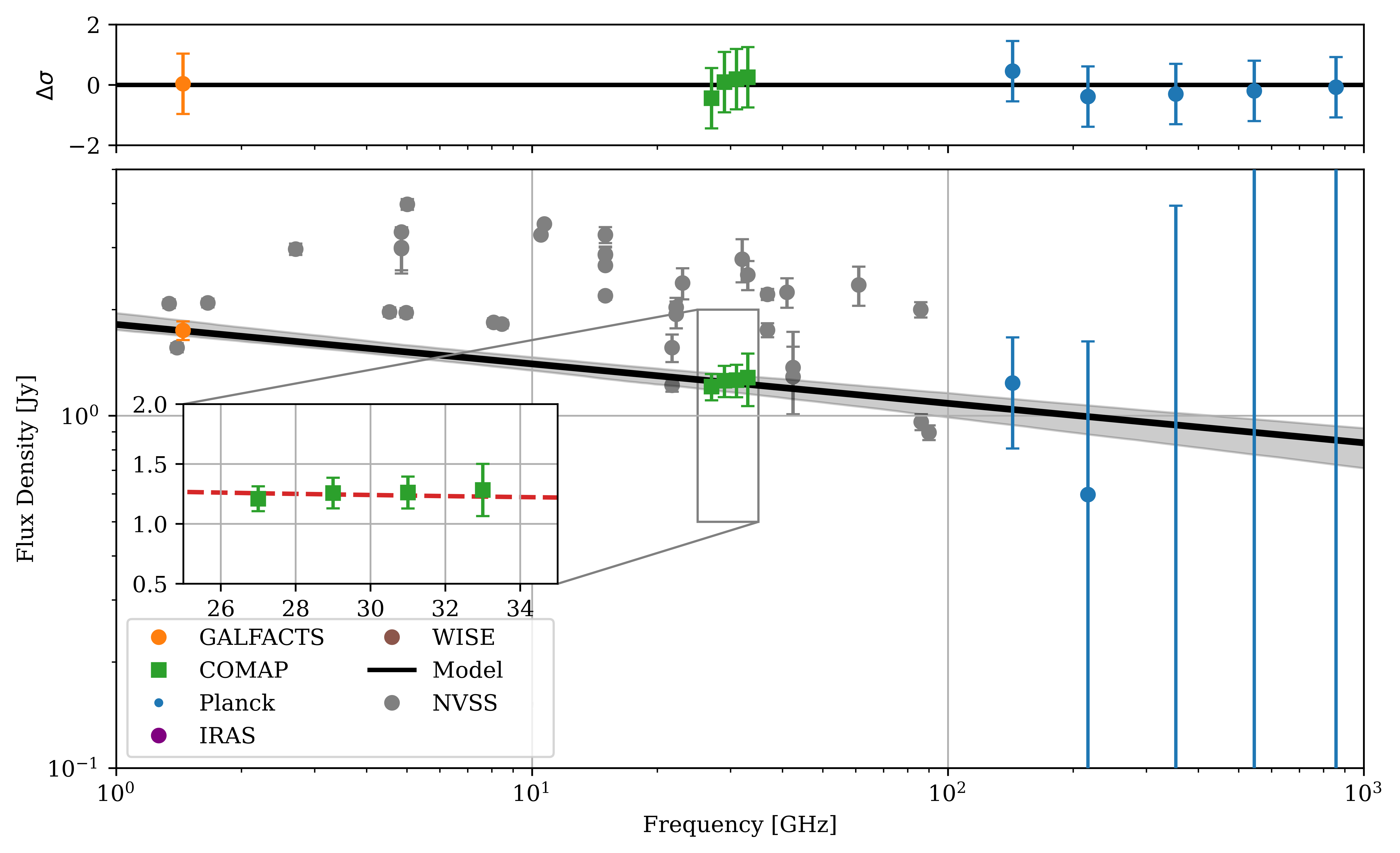}
    \caption{Best-fit power law spectrum to QSO~J0530+13 (NVSS~J053056+133155) as measured in the COMAP and GALFACTS maps. The \textit{grey} points are ancillary data from the literature taken from the NED extragalactic database, indicating significant time variability. Ancillary data from: \citet{Shimmins1975,Kuehr1981,Bennett1986,Wright1990,Becker1991,Gregory1991,Wiren1992,White1992a,Laurent-Muehleisen1997,Condon1998,Stanghellini1998,Horiuchi2004,Kellermann2004,Zavala2004,Kovalev2005,Rickett2006,Cooper2007,Healey2007,Petrov2007,Cara2008,Dodson2008,Hovatta2008,Vollmer2008,Massardi2009,Abdo2010,Ackermann2011,Koay2011,Lister2011,Nieppola2011,Planck2011_XV,Linford2012,Richards2014,Lister2015,Jeong2017}.}\label{fig:nvss}   
\end{figure*}

\begin{table}
    \centering 
    \caption{Flux densities measured in the COMAP and GALFACTS maps using the aperture given in \autoref{tab:apertures}. }\label{tab:nvss}
    \begin{tabular}{cc}
        \hline
        Frequency & Flux Density\\
        GHz & Jy \\
        \midrule
        1.447 & $1.70 \pm 0.10$\\
        27 & $1.20 \pm 0.10$\\
        29 & $1.30 \pm 0.10$\\
        31 & $1.30 \pm 0.10$\\
        33 & $1.30 \pm 0.10$\\ \hline 
    \end{tabular}
\end{table}

\autoref{fig:nvss} shows the spectrum of a radio source QSO~J0530+13 that is near to the main cloud, but is just outside of the region shown in \autoref{fig:finder}. QSO~J0530+13 is a well known variable, flat-spectrum QSO at a redshift of $z=2.070$ \citep{Dodson2008}. We give the flux densities measured by COMAP and GALFACTS in \autoref{tab:nvss}. We find that the flux densities measured by COMAP are consistent with other measurements in the literature, albeit COMAP estimates are among the lower measurements for the source at 30\,GHz. The fitted spectral index in \autoref{fig:nvss} is $\alpha = -0.11 \pm 0.03$, however this clearly is not an accurate model of the true source spectrum as can be seen by the the other available literature values (\textit{grey} points) taken from the NASA/IPAC Extragalactic Database (NED\footnote{\url{https://ned.ipac.caltech.edu}}).

\bsp	
\label{lastpage}
\end{document}